\documentclass[10pt,twocolumn,twoside] {IEEEtran}
\usepackage[caption=false,font=footnotesize]{subfig}
\usepackage{graphicx}
\usepackage{booktabs}
\usepackage{amsmath}
\usepackage{amssymb}
\usepackage{midfloat}
\usepackage{bm}
\usepackage{color, cite}
\usepackage{algorithm}
\graphicspath{{figures/}}

\begin{document}
\title{Radar and Communication Co-existence: an Overview}
\author{Le Zheng, \emph{Member}, \emph{IEEE},  Marco Lops, \emph{Fellow}, \emph{IEEE}, Yonina C. Eldar, \emph{Fellow}, \emph{IEEE}, and Xiaodong Wang, \emph{Fellow}, \emph{IEEE}
\thanks{Le Zheng is with Electronics \& Safety, Aptiv, Agoura Hills, CA, 91301, e-mail: le.zheng.cn@gmail.com.

Marco Lops is with the Dipartimento di Ingegneria Elettrica e delle Tecnologie dell'Informazione (DIETI) University of Naples ``Federico II" Via Claudio 21, 80125 Naples (Italy), e-mail: lops@unina.it.

Yonina C. Eldar is with the Department of Electrical Engineering, Technion, Israel Institute of Technology, Haifa 32000, Israel.
		
Xiaodong Wang is with Electrical Engineering Department, Columbia University, New York, USA, 10027, e-mail: wangx@ee.columbia.edu.		
}
}
\maketitle

\vspace{-1cm}

\begin{abstract}

Increased amounts of bandwidth are required to guarantee both high-quality/high-rate wireless services (4G and 5G) and reliable sensing capabilities such as automotive radar, air traffic control, earth geophysical monitoring and security applications. Therefore, co-existence between radar and communication systems using overlapping bandwidths has been a primary investigation field in recent years. Various signal processing techniques such as interference mitigation, pre-coding or spatial separation, and waveform design allow both radar and communications to share the {spectrum}. This article reviews recent work on co-existence between radar and communication systems, including signal models, waveform design and signal processing techniques. Our goal is to survey contributions in this area in order to provide a primary starting point for new researchers interested in these problems.

\end{abstract}

\section{Introduction}

The use of radar has been widened to numerous civilian applications including traffic control, remote sensing, car cruise control and collision avoidance. On a parallel track, the quest for ever increasing rates in wireless communications has pushed the carrier frequencies towards bands traditionally assigned to radar systems. This, along with the need to limit electromagnetic pollution, results in the scenario of co-existing radar and communication systems \cite{griffiths2015radar,blunt2018radar}. Emerging technologies in this field rely on concepts such as passive sensing, waveform diversity, co-design and the so called ``bio-inspired" strategies, wherein each part of a given architecture is seen as a sub-system whose design choices must be negotiated with the other constituent subsystems. To this last philosophy belong the class of cognitive systems, which are in turn intimately linked to the concept of Bayesian learning as a means to facilitate and sometimes enable individual decision-making \cite{griffiths2015radar,li2016optimum,CME18}.

The last few years have seen the growth of vibrant industrial and academic interest towards the convergence of sensing and communication functions. This has been affirmed by the announcement of the Shared Spectrum Access for Radar and Comm (SSPARC) program by the Defense Advanced Research Projects Agency (DARPA) \cite{evans2016shared} and the demands of sensing and communication for self-driving cars \cite{patole2017automotive}. As a result, a number of studies have been conducted, based on a variety of scenarios, degrees of cooperation between the coexisting systems, and design strategies.

The goal of this paper is to review existing results in this context and define a {\em taxonomy} of the different philosophies proposed so far. Three major architectures for co-existence have been henceforth defined:
\begin{enumerate}
	\item[(a)] Co-existence in spectral overlap;
	\item[(b)] Co-existence via cognition;
	\item[(c)] Functional co-existence.
\end{enumerate}
{Category (a) includes} architectures wherein both radar and communication systems are equipped with active transmitters using the same frequency spectrum. Here, the major problem is to eliminate or mitigate mutual interference while guaranteeing satisfactory performance for both functions. Different degrees of cooperation between the active systems have been so far accounted for. {Absolute lack of cooperation is assumed. For example,  in \cite{deng2013interference,sanders2012analysis} where the inherent resilience of properly designed coherent Multiple-Input Multiple-Output (MIMO) radars to the interference is exploited and attention is paid to the performance of the radar system only.} A similar ``radar-centric" philosophy is adopted in \cite{aubry2014radar,aubry2015new}, wherein co-existing communication users are safeguarded by limiting the amount of interference produced by the radar on given sub-bandwidths. Symmetrically, {uncooperative} ``communication-centric" approaches have been \textcolor{blue}{suggested} in a number of more recent studies, wherein countermeasures against the radar-induced interference are taken either at the communication receiver \cite{zheng2018adaptive} or, in the presence of some prior information, directly at the transmitter \cite{Liu2018MIMO,tuninetti}.  

Cooperation between the active systems, possibly operating in full spectral overlap, in order to negotiate the respective transmit policies and adjust the corresponding detection/demodulation strategies is the idea underlying co-design, first introduced in \cite{li2016optimum}, and further developed in \cite{lipetropulu,clancy,qian2018joint,zheng2018joint,khawar2014spectrum,puglielli2016design}. In this approach, which we define {\em holistic}, the co-existing systems are seen as constituent parts of a whole, so that the degrees of freedom under the designer's control are both the waveform(s) transmitted by the sensing systems and the code-book(s) employed by the communication systems. {These are jointly optimized so as to guarantee that both the communication and the radar performance are satisfactory. Co-design allows taking into account in the transceiver design effects such} as the reverberation produced by the radar, due to clutter or targets moving in close proximity to the communication receiver, range ambiguities and (random) Doppler frequencies. It is important to underline that these schemes are heavily knowledge-based and rely on information exchange between the constituent systems: this presupposes, on one hand, the presence of a ``fusion center" accessible to both systems, and, on the other, the accessibility of a common database, wherein the basic channel parameters are made available. 

In dynamic scenarios co-design may greatly benefit from {\em cognitive} paradigms. Here channel state is learned through suitable algorithms, which is conducive to the philosophy of co-existence via channel sensing put forth in \cite{CME18} and, more generally, to category (b) of the classification above. In fact, category (b) comprises systems wherein spectral overlap between the communication and radar transmitters is avoided through {\em cognition}, so that the corresponding channels are interference-free. Starting from the idea, proposed in \cite{lipetropulu} and borrowed from cognitive radio networks, of using pilot signals to estimate the channels and share the channel information between the subsystems, new approaches have been recently proposed wherein the radar and/or the communication system are able to ``learn" the environment without transmitting pilots or avoiding the need for coordination \cite{mishali2011xampling,mishali2010theory,eldar2015sampling,cohen2018analog}. In \cite{CME18}, for example, the SpeCX system combines sub-Nyquist multi-band sensing with sub-Nyquist radar \cite{cohen2018sub} so as to enable the radar to sense the communication channel at very low rates.

Category (c) comprises architectures wherein there is only one active transmitter, whereby co-existence is {\em functional}, but no interference is produced and no real resource negotiation takes place. Dual Function Radar Communication (DFRC) systems rely on combining radar and communication transmitters in the same hardware platform, which should be designed so as to guarantee the performances of both systems: the information is embedded \cite{blunt,hassanien2016dual,himedembed,8386661} in the radar signal, and a MIMO radar transmitter uses a combination of beam-forming and waveform diversity in order to direct information bits towards multiple communication receivers, without affecting the performance of the sensing function, while guaranteeing satisfactory Bit Error Rate (BER) performance. Opportunistic sensing systems, instead, consist of a receiver co-located with the communication transmitter and a dedicated software chain aimed at processing the received signal. The receiver can avail itself of some side information, such as timing and  transmitted data. This architecture has been proposed and theoretically assessed with reference to the 802.11ad format used in conjuction with a sensing system in an automotive environment \cite{Heath,802.11ad}. Passive radar systems also can be thought of as belonging to category (c) since they exploit other transmissions (communications, broadcast, or radionavigation) rather than having their own dedicated radar transmitter \cite{griffiths2015radar,griffiths2009passive}.

The article structure reflects the above categorization: in Sec. II we review systems in spectral overlap, by considering the cases of {uncoordinated} and coordinated transmission. Sec. III is devoted to cognition-based systems and Sec. IV focuses on functional duality. Concluding remarks and suggestions for future investigations are presented in Sec. V.

\section{Co-existence in spectral overlap}
\subsection{System Model}

\begin{figure}[htbp]
	\centering
	\subfloat{\includegraphics[width=3in]{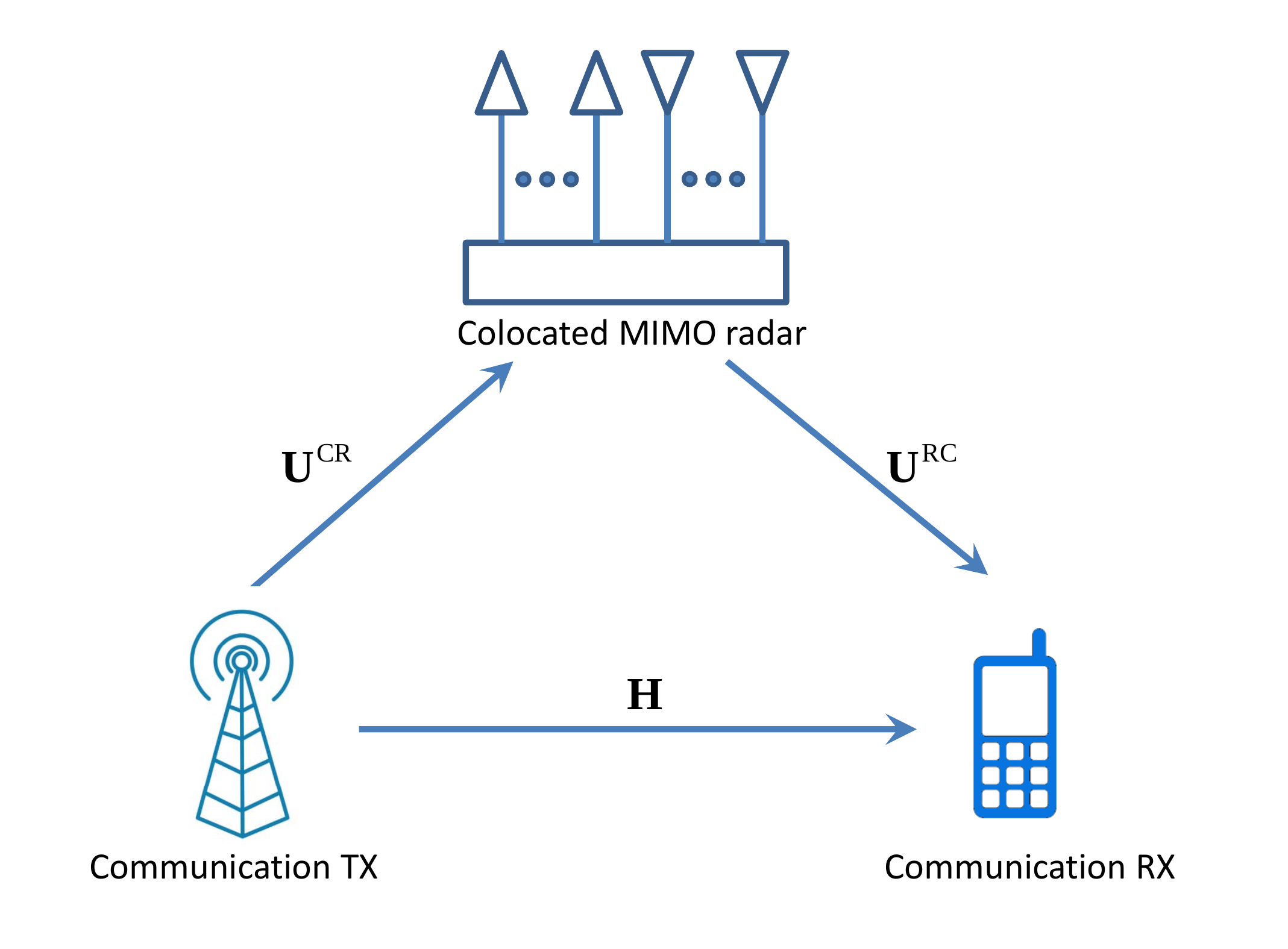}}
	\caption{{MIMO Communication system sharing spectrum with a MIMO radar system.}}
	\label{1}
\end{figure}

In the discussion below, we unify the Single-Input Single-Output (SISO) and MIMO settings as they are amenable to similar approaches. Thus, to keep the discussion as general as possible, we consider a scenario wherein a MIMO radar with $M_T$ transmit and $M_R$ receive (typically, but not necessarily co-located) antennas should co-exist with a MIMO communication system equipped with $N_T$ transmit and $N_R$ receive antennas, respectively, as illustrated in \figurename{1}. 

The MIMO radar transmits $M_T$ signals, where the signal transmitted from the $i$-th transmit element is characterized by a fast-time code $\mathbf c_i=[c_i(0), \ldots , c_i(P_r-1)] \in \mathbb{C}^{P_r}$. {The continuous-time waveform for the $i$-th transmit element is then given by}
\begin{equation}
\label{eq:basic-pulse}
\tilde c_i(t)=\sum_{p=0}^{P_r-1}c_i(p)\psi(t-p T_r). 
\end{equation}
Here $\psi(\cdot)$ is a Nyquist waveform of bandwidth\footnote{{Nyquist waveforms with bandwidth $B=\frac{1}{T_r}$ are strictly  band-limited, and therefore not time-limited. In practice, they are generated by truncation of an ideal waveform, whereby the discretization may incur some degree of aliasing: however, by allowing some excess bandwidth, this effect can be kept under control. A detailed discussion can be found in \cite{blunt2016overview}.}} $B=\frac{1}{T_r}$, i.e., such that its autocorrelation $R_{\psi}(\cdot)$ satisfies the condition $R_{\psi}(kT_r)=\delta(k)$, with $\delta (\cdot)$ denoting the Kronecker delta, and {$\frac{1}{T_r}$ is the fast-time coding rate}. 
{The product between the bandwidth and the ``effective" duration of these coded pulses is typically much larger than one. Therefore, these signals are sometimes referred to as ``sophisticated waveforms", as opposed to conventional un-sophisticated signals whose bandwidth is on the order of the inverse of their duration}.

In this architecture of Figure \ref{1}, every radiating element is allowed to transmit a train of $N$ {coded pulses of the form \eqref{eq:basic-pulse}}, spaced the pulse repetition time $T$ apart, and amplitude-modulated by a slow-time code ${\mathbf g}_i = [g_i(0), g_i(1), ..., g_i(N-1)]^T \in {\mathbb C}^{N}$. Thus the $i$-th element transmits the signal
\begin{equation}
s_i(t)=\sum_{n=0}^{N-1}g_i(n) {\tilde c}_i(t-nT). \label{eq:transmit-waveform}
\end{equation}

{Some special cases of the radar signal model \eqref{eq:transmit-waveform}} are as follows:
\begin{description}
	\item[1)] A single-antenna transmitter using a single signal with fast-time code $\mathbf c=[c(0), \ldots , c(P_r-1)]^T$, corresponding to $N=M_T=1$.
	\item[2)] A single-antenna transmitter using an amplitude-modulated train of pulses, corresponding to $M_T=1$, $P_r=1$. The train is uniquely determined by the slow-time code $\mathbf g = [g(0), \ldots , g(N-1)]^T \in \mathbb{C}^N$. {The} usual pulsed-radar corresponds to an all-one slow-time code.
	\item[3)] A multi-antenna transmitter wherein each antenna transmits a single sophisticated signal. As a consequence, $N=1$, $s_i(t)=c_i(t)$ and the $P_r \times M_T$ space-time code matrix $\mathbf C= \left[\mathbf c_1, \ldots , \mathbf c_{M_T}\right]$ is the degree of freedom to be employed at the transmitter side \cite{DeMaioLops}. 
	\item[4)] A multi-antenna transmitter wherein each antenna transmits a train of unsophisticated signals, amplitude-modulated by the slow-time code. In this case, $P_r = 1$ and the $N \times M_T$ space-time code matrix $\mathbf G= \left[\mathbf g_1, \ldots , \mathbf g_{M_T}\right]$ is the degree of freedom at the transmitter side \cite{zheng2018joint}. 
\end{description}
Radars use radio waves to determine the range, angle, or velocity of objects. {The operation of a typical MIMO radar receive chain is summarized in the box of Page 4.} The radar range resolution is dictated, for a given Signal-to-Noise Ratio (SNR), by the transmit bandwidth, i.e., $1/T_r$ in \eqref{eq:basic-pulse}. The velocity resolution is determined by the duration of coherent integration, i.e., $NT$ in \eqref{eq:transmit-waveform}. In situations 1) and 3) no Doppler processing is undertaken, mainly due to the fact that typical single-pulse durations are too short to allow measuring the Doppler shift induced by targets in moderate radial motion. In settings 2) and 4) moving objects generate steering vectors and Doppler shifts up to $\frac{1}{T}$ can be unambiguously measured. Likewise, pulse trains with Pulse Repetition Time (PRT) $T$ generate range ambiguities whereby scatterers located at distances corresponding to delays which are integer multiples of $T$ contribute to the same range cell.

\begin{figure*}
\newlength{\mylength}
\[
\setlength{\fboxsep}{15pt}
\setlength{\mylength}{\linewidth}
\addtolength{\mylength}{-2\fboxsep}
\addtolength{\mylength}{-2\fboxrule}
\fbox{%
	\parbox{\mylength}{
		\setlength{\abovedisplayskip}{0pt}
		\setlength{\belowdisplayskip}{0pt}

		Classic collocated MIMO radar processing traditionally includes the following stages:
		\begin{enumerate}
		\item[1)] Sampling: At each radar RX $1 \leq j \leq M_R$, the signal $r_j(t)$ is projected onto the orthonormal system $\{\psi(t-mT_r) \}_{m=0}^{P_r-1}$ and sampled at its Nyquist rate $B=\frac{1}{T_r}$, creating the samples $r_j(m)$, $0 \leq m \leq P_r-1$.
		 
		\item[2)] Matched filter: The sampled signal is convolved with the transmitted radar codes $\mathbf c_i$, $1 \leq i \leq M_T$. The time resolution attained in this step is $1/B$.
		
		\item[3)] Beamforming: The correlations between the observation	vectors from the previous step and the steering vectors corresponding to each azimuth on the grid defined by the array aperture are computed. 
		
		\item[4)] Doppler detection: The correlations between the resulting	vectors and Doppler vectors, with Doppler frequencies lying on the grid defined by the number of pulses, are computed. The Doppler resolution is $1/NT$.
		
		\item[5)] Peak detection: A heuristic detection process, in which knowledge of the number of targets, targets' powers, clutter location, and so on, may help in discovering targets' positions. For example, if we know there are $\kappa$ targets, then we can choose the $\kappa$-strongest points in the map. Alternatively, constant false alarm (FA) rate detectors determine {a} power threshold, above which a peak is considered to originate from a target so that a required probability of FA is achieved.
		 
		\end{enumerate}
		
		}}
\]
\end{figure*}

The signal model for the communication system is simpler in that we just have to distinguish between the case of single and multiple transmit antennas. In particular, we assume that the communication system operates on the same frequency band as the radar, occupying a fraction $\frac{B}{L}$ of its dedicated bandwidth. Setting $T_c=L/B$, the signal radiated by the $i$-th transmit element is written as
\begin{equation}
\label{eq:xi}
x_i(t)=\sum_{p=-\infty}^\infty {v}_i(p)\psi_L(t-pT_c),
\end{equation}    
where ${v}_i(p)$ is the data sequence to be transmitted, and $\psi_L(\cdot)$ satisfies the Nyquist criterion with respect to $T_c=LT_r$. The situation of full spectral overlap corresponds to $L=1$. We note that there may be a multiplicity of  narrow-band communication systems, each occupying a fraction of the radar bandwidth. 

Assume that the radar and the communication receivers are equipped with $M_R$ and $N_R$ receive antennas, respectively. The signal at the $j$-th antenna of the radar receiver (RX) can be cast in the form
\begin{eqnarray}
\label{eq:rj}
r_j(t) &=& \sum_{i=1}^{M_T} a_{i,j} s_i(t - \tau_{i,j}) + \sum_{i=1}^{N_T} (u_{i,j}^{\rm CR} * x_i)(t) \nonumber \\
&& + \sum_{i=1}^{M_T} (a_{i,j}^{\rm I} * s_i)(t) + n_{{\rm R},j}(t), 
\end{eqnarray} 
where $a_{i,j}$ is the target complex backscattering coefficient, including the path loss and the phase shift due to the target angle and position with respect to the transmit and receive antennas; ${u}_{i,j}^{\rm CR}(t)$ is the response of the channel from the communication Transmitter (TX) to the radar RX; $\tau_{i,j}$ is the delay of the target from the $i$-th TX to the $j$-th RX; ${a}_{i,j}^{\rm I}(t)$ is the response of the clutters; $*$ is the convolution operation; and $n_{{\rm R},j}(t)$ denotes the noise at the $j$-th RX antenna. Likewise, the signal received at the $j$-th antenna of the communication RX is given by
\begin{eqnarray}
\label{eq:yj}
y_j(t) = \sum_{i=1}^{N_T} ({h}_{i,j} * x_i)(t) + \sum_{i=1}^{M_T} ({u}_{i,j}^{\rm RC} * s_i)(t) + n_{{\rm C},j}(t), 
\end{eqnarray} 
where $h_{i,j}(t)$ is the channel response from the $i$-th communication TX to the $j$-th communication RX; $u_{i,j}^{\rm RC}(t)$ is the response of the interfering channel from radar TX to communication RX; and $n_{{\rm C},j}(t)$ denotes the noise of the $j$-th communication RX antenna. 

In \eqref{eq:rj}, the transmitted signal $s_i(t)$ is known and $u_{i,j}^{\rm CR}(t)$ can be estimated via pilot training. On the other hand, $x_i(t)$ and $a_{i,j}^{\rm I}(t)$ are unknown at the radar RX. The radar needs to detect the presence of the target, i.e., $a_{i,j} = 0$ for ${\cal H}_0$ and $a_{i,j} \neq 0$ for ${\cal H}_1$, and estimate the paramters $\tau_{i,j}$ and $a_{i,j}^{\rm I}(t)$. For the communication system given by \eqref{eq:yj}, ${h}_{i,j}(t)$ can be estimated via pilot training. In coordinated architectures, where the radar transmits pilots and communicates with the communication RX, ${u}_{i,j}^{\rm RC}$ and $s_i(t)$ are known at the communication RX, while in uncoordinated scenarios ${u}_{i,j}^{\rm RC}$ and $s_i(t)$ are both unknown. 
%The purpose here is to estimate $x_i(t)$ from the received signal.

Based on the models \eqref{eq:rj} and \eqref{eq:yj}, different co-existence scenarios can be analyzed. Sec. II-B discusses a radar-centric approach wherein a single-antenna radar transmits a single sophisticated signal with fast-time code, i.e. situation 1). Sec. II-C reviews some communication-centric approaches, assuming different degrees of prior knowledge as to the radar interference (i.e., scenarios 2) and 3) ). Sec. II-D focuses on coordinated design of the radar waveform(s) and the communication code-books, assuming the most general scenario (i.e., scenarios 3) and 4) ) of multiple transmit and receive antennas for both systems, with either slow-time or fast-time coding.

\subsection{Uncoordinated design: radar centric}

We begin by discussing a ``radar-centric" approach in which the radar function is considered {\em primary}, while unlicensed users are allowed to transmit in partial spectral overlap on the same bandwidth. Following \cite{aubry2014radar,aubry2015new}, we assume $N_I$ interferers of the form \eqref{eq:xi}. Their presence is acknowledged by limiting the amount of interference the radar produces on the shared bandwidths. The focus is on the design of the radar system, assumed to employ a single coded pulse according to situation 1) of the previous section, designed so as to guarantee the maximum possible Signal to Interference-plus-Noise Ratio (SINR) at the radar RX.

Assume that the radar RX is equipped with a single antenna and the interference is dominated by the direct path between the radar and the communication: the subscript $j$ can thus be removed from the variables in \eqref{eq:rj}. Thus, $r_j(t)$ becomes $r(t)$, and $u_{i}^{\rm CR}(t) = \delta(t - \tau_{i}^{\rm RC})$, with $\tau_{i}^{\rm RC}$ dictated by the distance between the $i$-th communication TX and the radar RX. Such a model holds for narrowband systems where the flat fading assumption is valid \cite{lipetropulu}, and can be extended to more sophisticated situations by using different forms of channel responses \cite{aubry2013knowledge}. For simplicity, we assume there is only one target and let the target delay {be} $\tau = 0$. 

Plugging \eqref{eq:transmit-waveform} into \eqref{eq:rj} and projecting the equation onto the orthonormal system $\{\psi(t-mT_r) \}_{m=0}^{P_r-1}$ leads to
\begin{eqnarray}
r(m) &=& \langle r(t) , \psi(t-mT_r) \rangle \nonumber \\
&=& \langle a \sum_{p=0}^{P_r-1}c_i(p)\psi(t-p T_r), \psi(t - mT_r) \rangle \nonumber \\
&& + \sum_{k=1}^{N_T} u_k \underbrace{ \langle x(t - \tau_k^{\rm CR}) , \psi(t - mT_r) \rangle}_{x_k(m)} \nonumber \\
&& + \underbrace{\langle \sum_{i=1}^{M_T} (a_{i}^{\rm I} * s_i)(t) , \psi(t - mT_r) \rangle}_{n_{\rm I}(m)} \nonumber \\
&& + \underbrace{\langle n_{\rm R}(t) , \psi(t - mT_r) \rangle}_{n_{\rm R}(m)}
\end{eqnarray}
with $a$ the target complex backscattering coefficient, including the path loss, and $u_k$ the coefficient of the interfering channel for user $k$. Denoting ${\mathbf r} = [r(0),r(1),...,r(P_r-1)]^T$, we have
\begin{eqnarray}
\label{eq:rec-sig}
\mathbf r = a \mathbf c + \sum_{k=1}^{N_I} u_k \mathbf x_k + \mathbf n_{\rm I} + \mathbf n_{\rm R} \in \mathbb{C}^{P_r},
\end{eqnarray}
with $\mathbf x_k = [x_k(0), x_k(1), ..., x_k(P_r-1)]^T$ the $k$-th communication user occupying the bandwidth, $\mathbf n_{\rm I} = [n_{\rm I}(0), n_{\rm I}(1), ..., n_{\rm I}(P_r-1)]^T \in \mathbb{C}^{P_r}$ the clutter, and $\mathbf n_{\rm R} = [n_{\rm R}(0), n_{\rm R}(1), ..., n_{\rm R}(P_r-1)]^T \in \mathbb{C}^{P_r}$ the noise term. 

Equation \eqref{eq:rec-sig} describes the model for the signal in the radar RX. Next, we discuss the interference from the radar to the communication users, i.e., the second term in \eqref{eq:yj}. As to the communication users coexisting with the radar of interest, we suppose that each of them is operating over a frequency band $[f_1^k,f_2^k]$, where $f_1^k$ and $f_2^k$ denote the lower and upper normalized frequencies for the $k$-th system, respectively. Following (2) and (3) in \cite{aubry2014radar}, the interfering energy produced on the $k$-th communication user is given by ${\mathbf c}^H {\mathbf R}_k {\mathbf c}$ where \color{black}
\begin{eqnarray}
\mathbf{R}_k (m,n) = \left\{ \begin{array}{l}
f_2^k - f_1^k ,{\text{ if }}m = n\\
\frac{{{e^{j2\pi f_2^k (m - n)}} - {e^{j2\pi f_1^k (m - n)}}}}{{j2\pi (m - n)}},{\text{ if }}m \ne n
\end{array} \right. \\
(m,n) \in \left\{1,2,...,P_r \right\}^2 \nonumber.
\end{eqnarray}
The covariance matrix $\mathbf M$ of the exogenous interference, i.e. of the signal-independent component of the overall interference $\sum_{k=1}^{N_I} u_k \mathbf x_k + \mathbf n_{\rm R}$, is assumed to be known or perfectly estimated. 

{The objective thus becomes to design the radar code $\mathbf c$ so as to maximize the SINR at the radar RX while ensuring that the interference produced on the co-existing communication users is smaller than a constrained value. Additional constraints to be enforced are an energy constraint on the radar code $\mathbf c$, and its ``closeness" to some reference code ${\mathbf c}_0$ with prescribed correlation properties \cite{aubry2014radar,aubry2015new}: the latter is also referred to as a ``similarity constraint". The design then reduces to solving the following constrained maximization problem:}
\begin{eqnarray} \label{eq:optimization-radar-centric}
&&\max_{\mathbf{c} \in {\mathbb C}^{N \times 1}} {\rm SINR} = a^2 \mathbf{c}^H \mathbf{M}^{-1} \mathbf{c} \\
&& {\rm s.t.} ~~~~ \sum_{k=1}^{N_R} \omega_k \mathbf{c}^{H} \mathbf{R}_k \mathbf{c} \leq E_{\rm I}, \nonumber \\
&& ~~~~~~~~  (1- \eta) \rho \leq \mathbf{c}^{H} \mathbf{c} \leq \rho,  \nonumber \\
&& ~~~~~~~~  \| {\mathbf c} - {\mathbf c}_0 \|_2 \leq \epsilon, \nonumber
\end{eqnarray}
{In the above equation,  the terms ${\mathbf c}^H {\mathbf R}_k {\mathbf c}$ represent the interference produced onto the $k$ communication receiver, $k= 1, 2, \ldots , N_R$, $E_{\rm I}$ the maximum interference that can be tolerated by the coexisting communication networks, $\omega_k \geq 0$ for $k = 1,2,...,N_R$ are weights that can be assigned to the coexisting wireless users based, for instance, on their distance from the radar and their tactical importance, $0 \leq \eta \leq 1$ is a design parameter which introduces some tolerance on the nominal interference level,  $\rho$ is the transmit energy of the radar. With relaxation, the optimization problem \eqref{eq:optimization-radar-centric} can be transformed into a convex optimization amenable to Semi-Definite Programming (SDP), which entails polynomial computational complexity \cite{aubry2015new}.} 

{The scenario leading to problem \eqref{eq:optimization-radar-centric} holds true only when the clutter is either absent or has rank one covariance matrix, i.e. is modeled as a specular image of the transmitted signal reflected towards the receiver by a point-like scatterer.} If, conversely, more complex channel models are considered, and the clutter covariance has rank larger than one (i.e., the point-like model does not carry over to reverberation), then constrained maximization of the SINR results in a fractional non-convex problem \cite{qian2018joint}.

\subsection{Uncoordinated design: communication centric}

The approach of optimizing radar waveforms, although theoretically well established, is not always applicable, mainly due to the fact that governmental and military agencies are unwilling to make major changes in their radar deployments, which may impose huge costs.  Thus, coexisting communication systems must be equipped with proper counter-measures to guarantee required Quality of Service (QoS) when the radar system(s) do not modify their transmission policy. Attention is thus shifted back to the communication transceiver, which explains the name ``communication-centric" design. The approaches so far available in the literature focus either on the receiver \cite{zheng2018adaptive}, when prior information on the radar signals is not available, or on the transmitter \cite{tuninetti}, when the structure of the radar transmitted waveform is known. 

Assume first the scenario considered in \cite{zheng2018adaptive}, wherein a multiplicity of radars may be potentially active in full spectral overlap with a communication system. Each radar is allowed to transmit a sophisticated waveform, but no prior knowledge as to the number of active systems, their distance from the communication receiver or the channel gains is available. The scenario is thus akin to the one outlined in situation 3) of the list of Sec. II-A, wherein $M_T$ now plays the role of the maximum number of potentially active emitters. The antennas of such a ``multiple input" system are widely spaced, so that the delays with which their signals arrive at the communication receiver are all different and unknown.

As to the communication signal, the scenario assumed in \cite{zheng2018adaptive} is fairly general. The transmitted symbols are assumed to undergo suitable pre-coding, where the choice of the pre-coding matrix dictates the type of system, ranging from Code-Division Multiple Access (CDMA) to Orthogonal Frequency-Division Multiplexing (OFDM). In particular, suppose the communication and radar systems have the same bandwidth, i.e., $L = 1$, $T_c = T_r$ and $\psi_L(t) = \psi(t)$. The signal transmitted by the communication system in the interval $[0,P_rT_r]$ is assumed to have the form
\[
 x(t) = \sum_{p= 0}^{P_r - 1} v(p) \psi(t - pT_r).
\]
In the above equation, $\mathbf v = [v(0), \ldots, v(P_r-1)]^T \in \mathbb{C}^{P_r}$ is tied to a generic $P$-dimensional data vector $\mathbf b_0=[b_0(0), \ldots , b_0(P-1)]^T$ to be transmitted as $\mathbf v= \mathbf A \mathbf b_0$, with $\mathbf A \in \mathbb{C}^{P_r \times P}$ a suitable matrix. Relevant special cases of the above model are the OFDM transmission format, wherein $P_r=P$ and $\mathbf A$ takes on the form of an Inverse Discrete Fourier Transform (IDFT) matrix, and a CDMA system with $P$ active users, wherein $\mathbf A$ contains the users' signatures \cite{zheng2018adaptive}. Here, in order to keep the discussion simple, we confine our attention to the case of direct transmission of the constellation points in full spectral overlap, so that $P=P_r$, $\mathbf b_0=\mathbf b \in \mathbb{C}^{P_r}$, $\mathbf A=\mathbf I_{P_r}$ ($\mathbf I_{P_r}$ denotes the identity matrix of order $P_r$). 

Suppose a single antenna communication RX, and single-tap model for both communication and interference channels. It is also assumed that the (typically high-power) radar transmitter is not saturating the front-end of the communication receiver. The communication signal in \eqref{eq:yj} can thus be re-written as
\begin{eqnarray}
\label{eq:yt}
&& y(t) =  h \sum_{p= 0}^{P_r - 1} b(p) \psi(t - pT_r)  \nonumber \\
&& + \sum_{m = 1}^{M_T} \sum_{p=0}^{P_r-1} u_m c_m(p) \psi(t - pT_r - \tau_m) + n_{\rm C}(t).
\end{eqnarray}
Here a flat-fading channel is assumed for the communication network where $h$ is the channel coefficient, $\tau_m$ and $u_m$ denote the (unknown) delay and complex coupling coefficient for the $m$-th radar, respectively. When $u_m=0$, the $m-$th transmitter is idle. We also assume that in each frame $P_r$ symbols are transmitted and that the frame sychronization between the radar and communication is guaranteed, i.e., the communication system is made aware of the beginning of the radar train pulse. This is a low-rate information, which can be shared once and for all, and regularly updated to account for possible timing drifts.

The communication RX has to accomplish jointly the two tasks of interference estimation/removal and data demodulation. For interference removal, we need to estimate $\tau_m$ and $u_m c_m(p)$ so as to substract the second term from \eqref{eq:yt}. Obviously, data demodulation and interference estimation are inherently coupled. {In \cite{JSTSP}, an iterative procedure is proposed for joint data demodulation and interference estimation, and a direct demodulation function $\mathbf{\hat b}^{(0)} = \Psi(\{ y(t) \}_{0 \leq t \leq P_r T_r} )$ is used as the initial step. In a general uncoordinated scenario, the communication receiver may not know the exact form of the interfering radar signals, but only rely on a coarse information of the family they belong to. A viable means to account for this uncertainty is to assume} that $\mathbf{c}_m$ lives in a low-dimensional subspace of ${\mathbb C}^{P_r}$, spanned by the columns of a known $P_r \times K$ matrix $\bm{\Phi} = [{\bm \phi}_0, {\bm \phi}_1, ..., {\bm \phi}_{P_r - 1}]^T \in {\mathbb C}^{P_r \times K}$ with $K \ll P_r$, i.e., $\mathbf{c}_m = \bm{\Phi} {\bm \alpha}_m$ for some unknown $\bm \alpha_m \in {\mathbb C}^{K}$, tied to the corresponding minimal and maximum distances of all of the potential radar transmitters from the receiver. 

Following \cite{zheng2018adaptive}, the signal $z^{(\ell)}(t) = y(t) - h \sum_{p= 0}^{P_r - 1} \hat b^{(\ell)}(p)\psi(t - pT_r)$ contains the superposition of the residual communication signal (due to demodulation errors), the residual radar interference and noise. To understand the joint interference removal and symbol demodulation algorithm proposed in \cite{zheng2018adaptive}, let us refer to the first iteration:
\begin{eqnarray}
\label{eq:z}
&& z^{(1)}(t) =  h \sum_{p= 0}^{P_r - 1} \underbrace{\left( b(p) - {\hat b}^{(0)}(p) \right)}_{\beta^{(0)}(p)}\psi(t - pT_r)  \nonumber \\
&& + \underbrace{\sum_{m = 1}^{M_T} \sum_{p=0}^{P_r-1} {\bm \phi}_m^T u_m \bm \alpha_m \psi(t - pT_r - \tau_m)}_{X(t)} + n_{\rm C}(t),
\end{eqnarray}
where $\tau_m$ for $m = 1,2,...,M_T$ are the desired unknown delays: in the above equation, the quantities ${\bm \phi}_m$ and $h$ are known, while the objects of interest to be estimated are $\tau_m$, $\beta^{(0)}(p)$ and $u_m \bm \alpha_m$.

\begin{figure}
	\centering
	
	\subfloat{\includegraphics[width=2.8in]{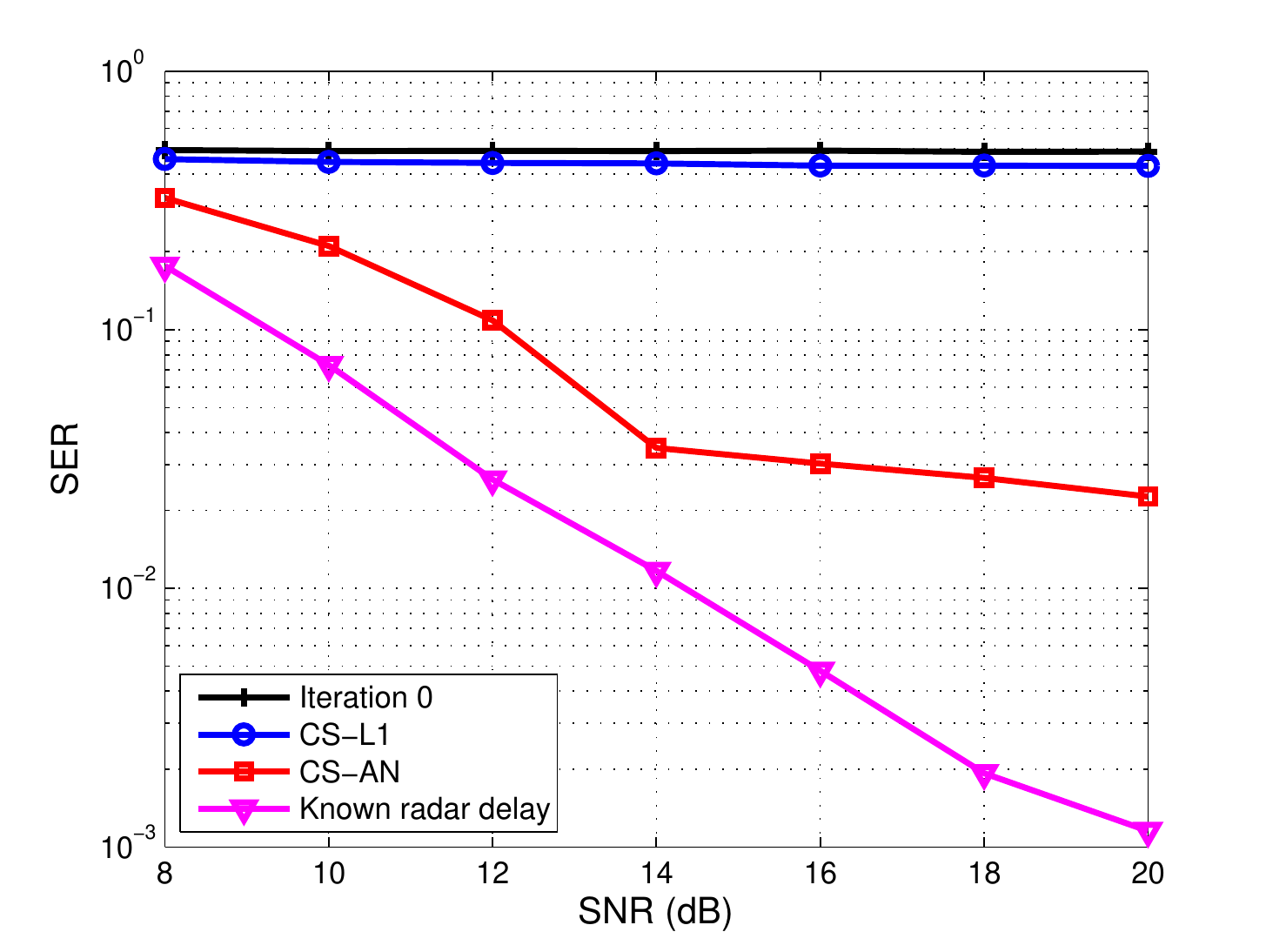}}
	
	\caption{Comparison of the algorithm SERs. Figure is taken from \cite{zheng2018adaptive}.}
	\label{fig:SER}
\end{figure}

Define $\bm \beta^{(\ell)} = [\beta^{(\ell)}(0), \beta^{(\ell)}(1), ..., \beta^{(\ell)}(P_r - 1)]^T$. Notice that building up iterations may rely on two types of sparsity, in that a) $X(t)$ in \eqref{eq:z} is a combination of at most $M_T$ components with unknown modulation $u_m \bm \alpha_m$, and $M_T \ll P_r$, and b) $\| \bm \beta^{(\ell)} \|_0$ has to be as small as possible. The problem can be solved by using the recently developed mathematical theory of continuous sparse recovery for super-resolution and in particular of Atomic-Norm (AN) minimization techniques \cite{zheng2018adaptive}. Figure \ref{fig:SER} illustrates the achievable results in terms of Symbol Error Rate (SER) for AN-based and CS-based methods, and allows assessing the loss due to the lack of prior knowledge as to the delays of the radar systems. 

A fairly different scenario is the one considered in \cite{tuninetti}, {where} it is assumed that the radar transmits a pulse train, possibly amplitude-modulated (according to the transmit format 2) of Sec. II-A). Perfect Chanel State Information (CSI) concerning the attenuation and delay of the radar signal in its travel to the communication receiver is assumed. Thus the interference generated by the radar onto the communication system is {\em intermittent}, and presents a large Peak-to-Average Power Ratio (PAPR), since it consists of pulses with large amplitudes. If the radar transmit code is a phase-only one (or if, more realistically, the pulse complex amplitudes vary significantly only in the phase), then a narrow-band communication system experiences an interference which is approximately a constant-envelope additive signal. Specifically, the interference is $({u}_{i,j}^{\rm RC} * s_i)(t) = \sqrt{I} e^{j \theta(t)}, t \in \Xi$ where $\theta(t)$ is the interference phase, assumed uniform in $[0, 2 \pi)$, $I = |ug|^2$ denotes the average power of the radar interference, assumed known, and $\Xi$ designates the time intervals where the communication system is interfered. The communication transmitter in turn randomly selects the symbols to be transmitted from the set ${\cal B}= \{\tilde b_1, \tilde b_2,...,\tilde b_Q \}$ of unit-energy and equally-likely points. Exploiting the statistical independence between these symbols and $\theta(t)$,  the optimal decoding regions can be obtained, and the constellation $\cal B$ can be designed to maximize the transmission rate and/or minimize the error rate.

\subsection{Coordinated design}

The major drawback of the previous approaches is that they rely on a simplified scenario wherein several important phenomena are not accounted for:
\begin{enumerate}
	\item[-] The radar system, especially when operating in search mode, generates reverberation from the surrounding environment, the so-called clutter, which impairs not only its own performance, but also the performance of the communication system.
	\item[-] The scattering centers generating clutter could have radial motion with respect to both the radar and the communication receivers, thus generating Doppler shifts that should be accounted for if slow-time coding is considered.
\end{enumerate}
Cooperation between the active systems, possibly operating in full spectral overlap, in order to negotiate the respective transmit policies and adjust the corresponding detection/demodulation strategies is the idea underlying co-design, first introduced in \cite{li2016optimum}, and further developed in \cite{lipetropulu,qian2018joint,zheng2018joint}. It is generally assumed that the radar and the communication system may exchange information. The availability of large data-bases accurately mapping the scattering characteristics of large areas has allowed the development of {\em cognitive systems} (see, e.g., \cite{DeMaiocognitive,Stoicacognitive}): joint design of the radar waveform(s) and the communication system codebook thus appears as a natural means to allow co-existence by preserving the performances of both. 
 
Consider an $N_T \times N_R$ communication system co-existing in full spectral overlap with an $M_T \times M_R$  MIMO radar with closely spaced antennas and co-located transmitter and receiver. We denote by $\mathbf D$  the space-time code matrix of the radar: if fast-time coding is adopted, then $\mathbf D = \mathbf C$, with $\mathbf C$ defined in situation 3). If, instead, slow-time coding is undertaken, then $\mathbf D = \mathbf G$ and situation 4) occurs. Denote by $\mathbf V$ the signal matrix of the communication system, composed of the $N_T$ spatial codewords emitted in successive epochs. Specifically, ${\mathbf V} = [{\mathbf v}(0), {\mathbf v}(1),...,{\mathbf v}(P_r-1)] \in {\mathbb C}^{N_T \times P_r}$ where ${\mathbf v}(p) = [{v}_{1}(p), {v}_{2}(p), ..., {v}_{N_T}(p)]^T$
is the spatial codeword transmitted at epoch $p$. % The returns from a target located in a given range-azimuth cell is embedded in
%\begin{itemize}
%	\item Receiver noise, which is modeled - as usual - as Gaussian and white;
%	\item Interference from the co-existing communication system, that we assume to operate in full bandwidth overlap;
%	\item Reverberation from the surrounding environment: this could be due to ambient scatterers (land, sea, vegetation) or to different targets moving across the scene.
% \end{itemize} 
Projecting the received signal \eqref{eq:rj} and \eqref{eq:yj} onto the orthonormal system $\{\psi(t-mT_r) \}_{m=0}^{P_r-1}$ leads to:
\begin{eqnarray}
\label{eq:mimo_radar}
\mathbf R &=& {\mathbf A} {\mathbf D} + {\mathbf U}^{\rm CR} {\mathbf V} + {\mathbf A}^{\rm I} {\mathbf D} + {\mathbf N}_{\rm R}, \\
\label{eq:mimo_comm}
\mathbf Y &=& {\mathbf H} {\mathbf V} + {\mathbf U}^{\rm RC} {\mathbf D} + {\mathbf N}_{\rm C},
\end{eqnarray}
where ${\mathbf A}\in {\mathbb C}^{M_R \times M_T}$ is the response of the target to be detected; ${\mathbf A}^{\rm I} \in {\mathbb C}^{M_R \times M_T}$ is the response of the clutters; ${\mathbf N}_{\rm R}$ is the noise at the radar RX; ${\mathbf N}_{\rm C}$ is the noise at the communication RX; ${\mathbf U}^{\rm CR} \in ^{M_R \times N_T}$ is the interfering channel from communication TX to radar RX; ${\mathbf U}^{\rm RC} \in ^{N_R \times M_T}$ is the interfering channel from communication TX to radar RX; ${\bf{H}} \in {\mathbb C}^{N_R \times N_T}$ is the channel matrix from communication TX to radar RX. {In \eqref{eq:mimo_comm} the MIMO communication system is assumed to have perfect channel state information - i.e. knowledge of ${\mathbf V}$ - to be periodically shared with the radar system through a dedicated channel.} 

In \eqref{eq:mimo_radar}, the purpose of the MIMO radar is to detect the presence of a target ($\mathbf A = \mathbf 0$ for ${\cal H}_0$ and $\mathbf A \neq \mathbf 0$ for ${\cal H}_1$) and estimate the matrix $\mathbf A$ which is related to the target parameters such as angle and velocity. An important additional degree of freedom is the space-time filter that can be applied to the radar signal $\mathbf R$ in \eqref{eq:mimo_radar}. Let $\mathbf{\tilde r} = {\rm vec}({\mathbf R}) = [{\mathbf r}(0)^T, {\mathbf r}(1)^T,...,{\mathbf r}(P_r-1)^T]^T$ with ${\mathbf r}(p)$ the $(p+1)$-th column of $\mathbf R$. The filtered signal becomes
\begin{eqnarray}
\label{eq:barr}
{\bar r} = \mathbf{\tilde w}^T \mathbf{\tilde r},
\end{eqnarray}
with $\mathbf{\tilde w}  \in {\mathbb C}^{M_R P_r \times 1}$. We recall here that the receive filter is of fundamental importance in coherent MIMO radar, since time filtering regulates the transmit beam-width, while space filtering controls the receive beam-pattern.

A possible criterion to exploit transmitter coordination for a coherent MIMO radar co-existing with a communication system is to force the radar waveforms $\mathbf D$ to live in the null space of the interference channel ${\mathbf U}^{\rm RC}$ via a spatial approach \cite{khawar2014spectrum}. The MIMO structure indeed provides the degrees of freedom to suitably design the space-time code matrix determining the probing signal. To illustrate further, assume that the model of situation 3) of Sec. II-A is in force, and that the fast-time space-time code matrix ${\bf C}$ is to be designed. To this end, we regroup the signals transmitted by the MIMO radar in the vectors ${\mathbf c}(p) = [c_1(p), c_2(p),...,c_{M_T}(p)]^T$, encapsulating the spatial codeword transmitted for the $p-$th sub-pulse. Consider the situation that $\bar N$ communication RX's exist, and let the interference channels of the communication RX's are $\{ {\mathbf U}^{(1)}, {\mathbf U}^{(2)},...,{\mathbf U}^{(\bar N)}\}$. In \cite{clancy}, where the idea is fully developed, these abstract ``communication RX's" are actually clusters of base-stations. The interference that would be produced onto the $n$-th communication RX is ${\mathbf U}^{(n)} {\mathbf c}(p)$. At the MIMO radar, the channel state information can be estimated using a blind null space learning algorithm \cite{noam2013blind}. 

Our goal here is to assure zero interference to one of the communication RXs with minimum degradation in the radar performance. Suppose we want no interference at the $n$-th communication RX. The communication signal can be projected onto the null space of the channel $\mathbf U^{(n)}$. The null space ${\cal N}({\mathbf U}^{(n)})= \{ {\mathbf c} \in {\mathbb C}^{M_T}: {\mathbf U}^{(n)}{\mathbf c} = {\mathbf 0} \}$ can then be calculated based on Singular Value Decomposition (SVD). Specifically, letting ${\mathbf U}^{(n)} = {\bm \Upsilon}_1 {\bm \Sigma} {\bm \Upsilon}_2^H$, the right singular vectors corresponding to vanishing singular values are collected in $\bm{\bar \Upsilon}_2$ for the formation of the projection matrix ${\mathbf P}_{\bm{\bar \Upsilon}_2}^{(n)} = \bm{\bar \Upsilon}_2 (\bm{\bar \Upsilon}_2^H \bm{\bar \Upsilon}_2)^{-1} \bm{\bar \Upsilon}_2^H $. The transmitted radar signal is thus the projection of ${\bf c}(p)$ onto the null space, i.e.,
\begin{eqnarray}
\label{eq:beamforming-signals}
\tilde{\mathbf c} (p) = {\mathbf P}_{\bm{\bar \Upsilon}_2}^{(n)} {\mathbf c} (p).
\end{eqnarray}
The precoder ${\mathbf P}_{\bm{\bar \Upsilon}_2}^{(n)}$ inevitably introduces correlation among the signals emitted by the different transmit elements,  thus generating some performance loss for target direction estimation. Note that the radar waveform is orthogonal to one communication channel, but not to all. The MIMO radar selects the best interference channel, defined as
\begin{eqnarray}
\mathbf U_{\rm Best} = \mathbf U^{(i_{\max})} , {\rm with}~~ i_{\max} = \arg \max_{1 \leq i\leq \bar N} {\rm dim} [{\cal N} (\mathbf U^{(i)})], 
\end{eqnarray}
and avoids the worst channel, defined as
\begin{eqnarray}
\mathbf U_{\rm Worst} = \mathbf U^{(i_{\min})} , {\rm with}~~ i_{\min} = \arg \min_{1 \leq i\leq \bar N} {\rm dim} [{\cal N} (\mathbf U^{(i)})].
\end{eqnarray}

% Selecting the best channel for radar's signal projection assures minimum degradation in the radar performance while assuring zero interference to one of the communication RXs. 
In general, in the fully cooperative scenario outlined in \cite{clancy} the radar can take a snapshot of the interference situation for each cluster, and broadcast it to allow proper user(s) assignment protocols. Users may then be assigned to less or more interfered base stations based on priority order. 

\begin{figure}
	\centering
	
	\subfloat{\includegraphics[width=2.8in]{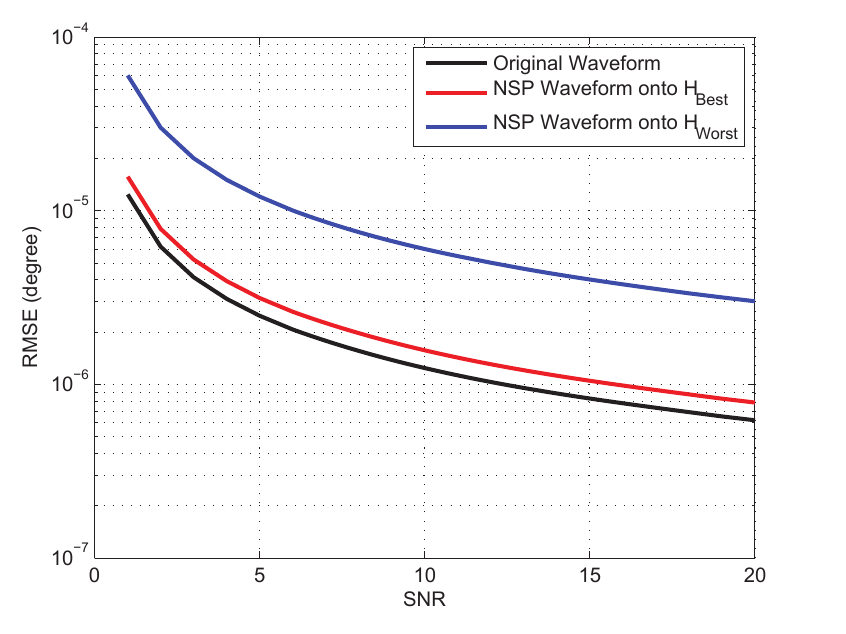}}
	
	\caption{Cramer Rao Bound (CRB) on target direction estimation RMSE as a function of the SNR, when $\mathbf U_{\rm Best}$ and $\mathbf U_{\rm Worst}$ (marked as $\mathbf H_{\rm Best}$ and $\mathbf H_{\rm Worst}$, respectively) channels are selected. Figure is taken from \cite{khawar2014spectrum}.}
	\label{fig:Clancy}
\end{figure}

{In Figure \ref{fig:Clancy}, we compare the root-mean-square-error (RMSE) of the target direction estimation under different radar waveforms. Note that the estimation performance as the null-space projection (NSP) waveform onto $\mathbf U_{\rm Best}$ is closer to the performance of the original radar waveform in RMSE sense. Thus, by an appropriate selection of the interference channel, the degradation in the radar performance, due to the NSP of the waveform, can be reduced.}  

\begin{figure}
	\centering
	
	\subfloat{\includegraphics[width=3.5in]{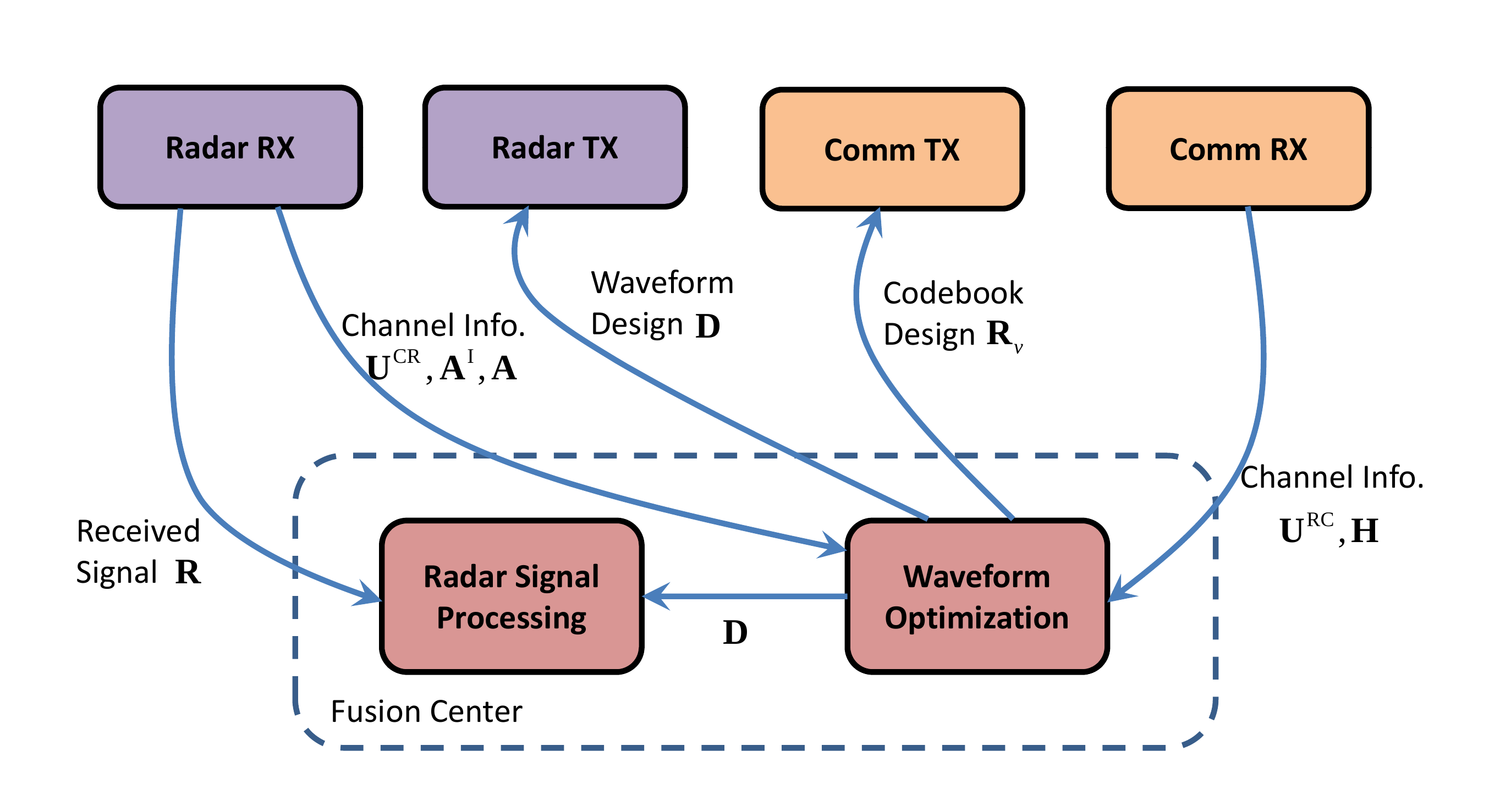}}
	
	\caption{Schematic structure of a coordinated design of radar and communication waveforms based on optimization.}
	\label{fig:Codesign}
\end{figure}

A MIMO radar can operate without creating interference at any of the communication RXs if the number of radar transmit antennas is greater than the sum of the requested degrees of freedom of all of the communication RXs \cite{babaei2013practical}. Cooperation between all of the BS’s and radar allows forming the interference matrix $\mathbf{\bar U} = \left[ \mathbf U^{(1) T}, \mathbf U^{(2) T},..., \mathbf U^{(\bar N) T} \right]^T \in {\mathbb C}^{\bar N N_R \times M_T}$. Applying the previous strategy yields $\mathbf{\tilde c} (p) \in {\cal N}( \mathbf{\bar U} )$. Other alternative strategies may rely on forcing the radar waveform to be designed according to an MMSE criterion (rather than to the aforementioned Zero-Forcing).

More general approaches for the coordinated design of radar and communication are based on optimization methods (illustrated in Figure \ref{fig:Codesign}). We assume that the radar uses an $P_r \times M_T$ space-time code matrix $\mathbf C$: the extension to the slow-time coding can be undertaken by changing the time scale, considering the Doppler effect in the signal model of \eqref{eq:mimo_radar} and \eqref{eq:mimo_comm}, and solving for the slow-time space-time matrix $\bm G$ \cite{qian2018joint}. The space-time filter $\mathbf{\tilde w}$ in \eqref{eq:barr} can also be optimized so as to improve radar performance.

Assume that the SINR is the figure of merit of interest to the radar and let $\cal Q$ be the figure of merit chosen for the communication system: they depend on $\mathbf D$, on the symbol matrix $\mathbf V$ (or on some statistical feature thereof if random coding is undertaken), as well as on a number of channel parameters tied to the reverberation, that we combine in an unspecified array $\mathbf Z$. A suitable figure of merit guaranteeing the performance of the communication system is the mutual information between the input symbol stream and the observations \cite{li2016optimum,qian2018joint}. In particular, the mutual information averaged over $P_r$ time slots, assuming Gaussian interference is
\begin{equation}
C = \frac{1}{P_r} \sum_{p=0}^{P_r-1} {\log_2 \det \left( {{\bf{I}}_{N_{R}L} + {\bf{R}}_{\rm Cin}^{-1}{\bf{{H}}}{{\bf{R}}_{v}(p)}{{\bf{{H}}}^H}} \right)},
\label{eq:capacity}
\end{equation}
where ${\mathbf R}_v(p) = {\mathbb E} \left[{\mathbf v}(p) {\mathbf v}(p)^H \right]$ is the covariance matrix of the communication codebook, and ${\bf{R}}_{\rm Cin} \in {\mathbb C}^{N_R \times N_R}$ is the covariance of interference plus noise, assumed either known or perfectly estimated. The transceivers are designed so as to guarantee prescribed QoS to both systems. 

A possible optimization problem can be formulated as:
\begin{equation}
\begin{array}{l} \label{eq:problem_full1}
{\cal P}\left\{ \begin{array}{l}
\mathop \text{max}\limits_{{{{\mathbf D, \{\mathbf R_v(p)\}, \mathbf{\tilde w}}} }}\:
\text{SINR}\left( {\bf D}, \{\mathbf R_v(p)\}, \bf{Z}, {\mathbf{\tilde w}} \right),  \\
{\rm{{s}}}{\rm{.t}}{\rm{.  }} ~~
{\cal Q}\left( \mathbf{D}, \{\mathbf R_v(p)\}, \mathbf{Z} \right) \geq {\cal Q}_0, ~~~~ \text{QoS of Comm. Syst.} \\
\ \ \quad g_i(\mathbf D) \leq 0, i=1,...,I_R, ~~~~~~ \text{Rad. wav. constr.} \\
\ \ \quad f_i(\{\mathbf R_v(p)\}) \leq 0, i=1,...,I_T, \text{Comm. codes. constr.}
\end{array} \right.
\end{array}
\end{equation}
where $\text{SINR}(\cdot)$ is the SINR at the output of the radar receiver, $g_i(\cdot)$ and $f_i(\cdot)$ are a set of constraints forced on the radar and communication transmitted signals, respectively. The problem in \eqref{eq:problem_full1} is typically non-convex, but Alternating Maximization (AM) techniques have been proposed and implemented in \cite{qian2018joint,lipetropulu} through decompositions into sub-problems which are either convex or solvable through fractional programming methods. In \cite{qian2018joint}, for example, problem \eqref{eq:problem_full1} has been reformulated for slow-time coding, explicitly accounting for Doppler shifts of both the target to be detected and the environmental reverberation.

%in terms of the vectors obtained by stacking up the spatial observations. Let $\mathbf{\tilde v} = {\rm vec}({\mathbf V})$ and define $\mathbf{\tilde R}_v = {\mathbb E} \left( \mathbf{\tilde v} \mathbf{\tilde v}^H \right)$. The optimized covariance can be obtained by reformulating \eqref{eq:problem_full1} with $\{\mathbf R_v(p)\}$ replaced by $\mathbf{\tilde R}_v$ and solving it.

\section{Co-existence via cognition}

\subsection{Environment sensing techniques}

The idea of knowledge-based design is central for spectrum-sharing systems \cite{CME18,deng2013interference,babaei2013practical,khawar2014spectrum,puglielli2016design,lipetropulu,zheng2018joint}.
%  exploiting the spatial degrees of freedom granted by a MIMO radar , adaptive transmit/receive strategies to test the occupancy of the frequency bands \cite{CME18} or joint design of radar waveform and communication system code-book given the interference channel estimation \cite{lipetropulu,zheng2018joint}. 
The communication and/or the radar system undertake suitable ``environment sensing" phases in order to determine the transmit policies. Inspired by cooperative methods in cognitive radio networks, \cite{lipetropulu} uses pilot signals to estimate the channels and feed back the channel information between the subsystems, possibly assigning to one of them a functional priority, like, e.g., in\cite{zhang2008exploiting,zhang2010dynamic}, where the radar is considered primary. These approaches rely on a centralized architecture, namely on a strict coordination between the active players in order to allow co-existence.

More recently, approaches wherein the radar and/or the communication system are able to ``learn" the environment without transmitting pilots or avoiding the need for coordination have been proposed. These advanced approaches are discussed in the following two scenarios:
\begin{itemize}
	\item[1)] Environment sensing at the communication RX: A communication system shares its spectrum with an ensemble of potential interferers, i.e., a set of radar/sensing systems. The interfering waveforms from the radars lie in the subspace of a known dictionary, and impinge on the communication RX with unknown, possibly time-varying delays and coupling coefficients. 
	
	\item[2)] Environment sensing at the radar RX: A sparse target scene is assumed, allowing to reduce the radar sampling rate without sacrificing delay and Doppler resolution. The Xampling framework can be adopted, where the system architecture is designed for sampling and processing of analog inputs at rates far below Nyquist, whose underlying structure can be modeled as a union of subspaces \cite{cohen2018sub}. 
	
\end{itemize}

\begin{figure*}[h]
	\centering
	\includegraphics[width=18cm]{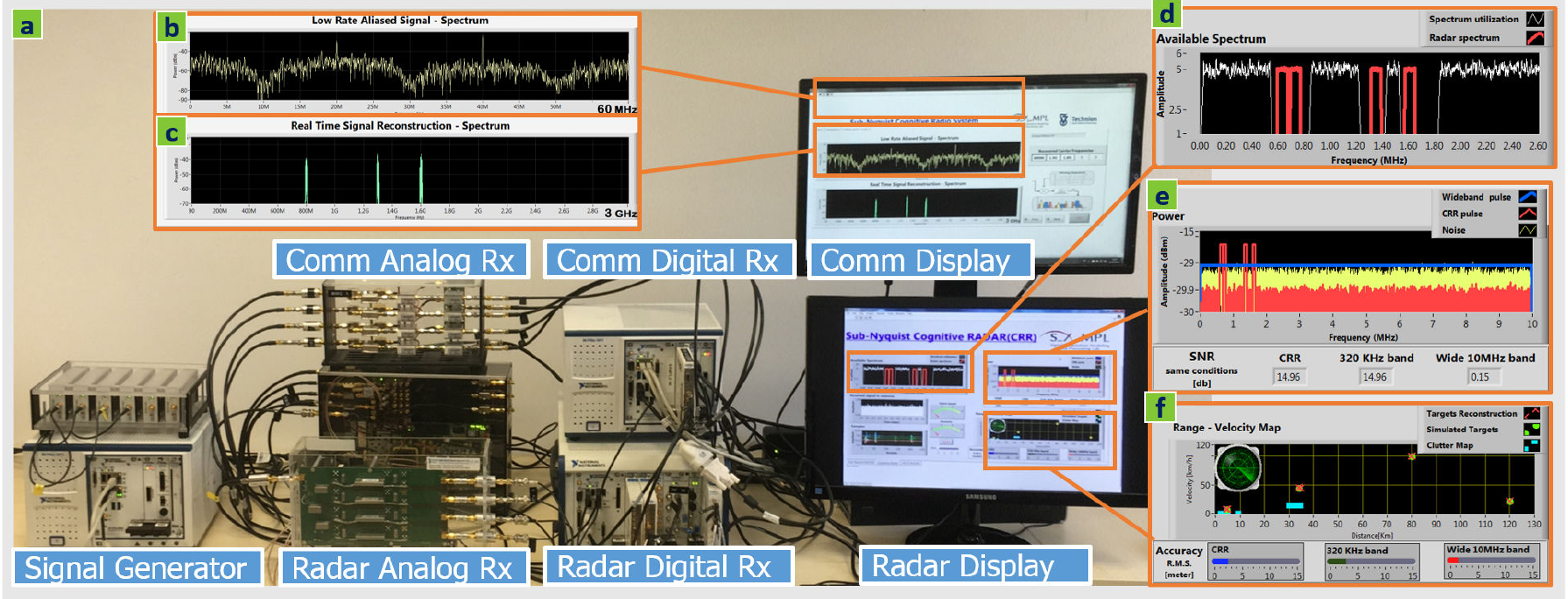}
	\caption{(a) SpeCX prototype. The system consists of a signal generator, a cognitive radio receiver based on the MWC, a communication digital receiver, and a cognitive radar. SpeCX comm system display showing (b) low rate samples acquired from one MWC channel at rate 120 MHz, and (c) digital reconstruction of the entire spectrum from sub-Nyquist samples. SpeCX radar display showing (d) coexisting communication and cognitive radar, (e) cognitive radar spectrum compared with the full-band radar, and (f) range-velocity display of detected and true locations of the targets. Figure is taken from \cite{CME18}.}
	\label{fig:SpeCX}
\end{figure*}

The former situation has been described in Sec. II-C. The communication RX must be made adaptive, in order to accomplish jointly the two tasks of interference estimation/removal and data demodulation. {For the latter situation, the SpeCX system (shown in Figure \ref{fig:SpeCX}) was proposed in \cite{CME18}, and combines sub-Nyquist multi-band sensing with sub-Nyquist radar so as to enable the radar to sense the communication channel at very low rates. Compared to the other works, SpeCX presents a complete solution that shows recovery of both the radar and communication signal with minimal known information about the spectrum.} 

More specifically, a sub-Nyquist cognitive radio is first implemented to sense the communication channel and determine which bands are occupied. This can be done using the modulated wideband converter (MWC), a sub-Nyquist communication receiver developed specifically for this task which is capable of detecting sparse signals at very low rates \cite{mishali2011xampling,mishali2010theory,eldar2015sampling,cohen2018analog}. Once the empty bands in the spectrum are identified, a cognitive radar receiver is employed which transmits a wideband signal that consists of several narrow band signals, in the vacant frequency bands \cite{cohen2016towards}. Using the radar Xampling paradigm, it can be shown that high resolution delay and Doppler can be performed from such a multiband wideband radar signal by combining methods of sub-Nyquist sampling and compressed beamforming \cite{eldar2015sampling,bar2014sub,baransky2014sub}. This allows to detect targets with high resolution while using a transmit signal that consists of several narrow bands spread over a wide frequency regime. The advantage of such a system is that the total bandwidth occupied is small while still allowing for high resolution. {This enables transmission of an adaptive radar signal that can co-exist with a standard communication channel and also leads to low rate, low power receivers.}

\subsection{Knowledge-based design}

% The transmitter of radar system senses the channel and avoids producing interference. 

In this subsection, we survey knowledge-based radar transmission designs based on environment sensing. For example, in some settings, the radar interference can be eliminated by forcing the radar waveforms to live in the null space of the interference channel between the radar transmitters and the communication receiver \cite{khawar2014spectrum}. This idea is well studied in the cognitive radio research community, and also applied to spectrum sharing systems. Typical approaches include exploiting the spatial degrees of freedom granted by a MIMO radar \cite{deng2013interference,babaei2013practical,khawar2014spectrum,puglielli2016design} and adaptive transmit/receive strategies to test the occupancy of the frequency bands \cite{CME18}.

In \cite{CME18,eldar2015sampling} the bands selected by the radar are chosen to optimize the radar probability of detection. More specifically, after the communication signal support is identified, denoted as ${\cal F}_c$, the communication receiver provides a spectral map of occupied bands to the radar. Equipped with the detected spectral map and known radio environment map (REM), denoted as ${\cal F}_r$, the objective of the radar is to identify an appropriate transmit frequency set that does not overlap with the union of ${\cal F}_c$ and ${\cal F}_r$, and maximizes the probability of correct detection. This probability increases with the SINR when the probability of false alarm is fixed. Therefore, it is proposed to maximize the SINR or minimize the spectral power in the undesired parts of the spectrum. This is achieved by using a structured sparsity framework \cite{huang2011learning}. Additional requirements of transmit energy constraints, range sidelobe levels, and minimum separation between the bands can also be imposed. Once the optimal radar support is identified, a suitable waveform code may be designed over this support. 

Another approach for waveform design is based on spectral notching that minimizes transmit energy in specific frequency bands, rather than designing a waveform that is avoiding interference, while maintaining desirable envelope and sidelobe characteristics \cite{frost2012sidelobe}. A waveform designed to avoid transmitting in specific bands, a spectrally-disjoint waveform, must be characterized using other metrics since interference is not driving the design, and thus, no such SINR can be calculated. Such metrics include average power levels in the undesired frequency bands, peak sidelobe levels, and integrated sidelobe levels.

%A different approach is to use the spectrally-disjoint waveform to avoid transmitting in specific bands. 

\section{Functional co-existence}

\subsection{Embedding data into radar waveforms}

\begin{figure}[htbp]
	\centering
	\subfloat{\includegraphics[width=2.6in]{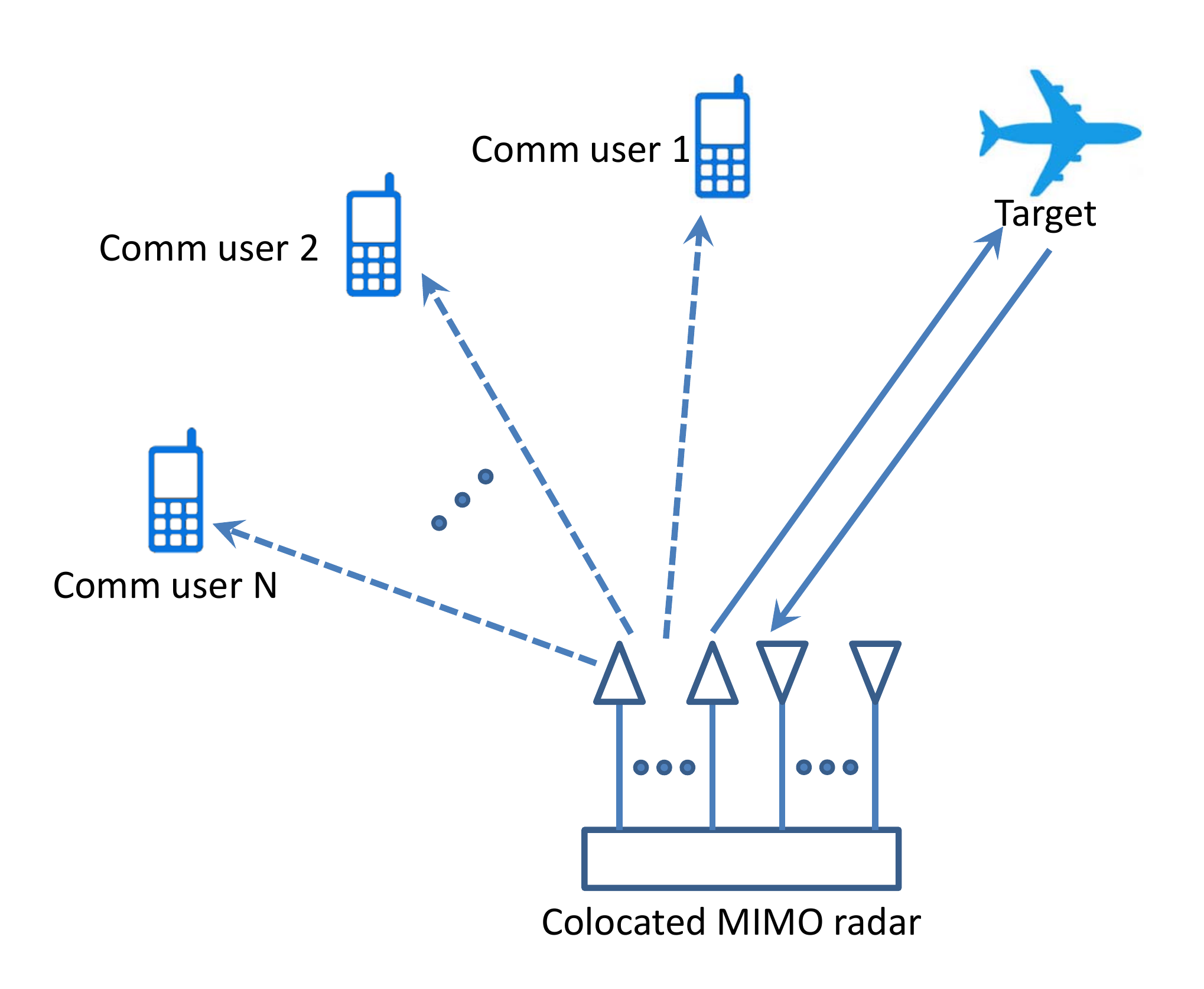}}
	\caption{{Dual-function radar communications.}}
	\label{fig:DFRC}
\end{figure}

A fairly natural evolution of radar and communication co-existence is to use radar to perform communication, also known as Dual Function Radar Communication (DFRC)\cite{ahmad}. This approach is illustrated in Figure \ref{fig:DFRC}, wherein radar and communication systems are combined in the same hardware platform, usually with the same waveform or transmitter, which should be designed so as to guarantee the performance of both systems. As echoed by the name itself, in these architectures co-existence is basically ``functional" and no spectrum overlap or resource negotiation takes place. This philosophy relies on the strategy of ``information embedding". Consider a joint radar communication platform equipped with $M_T$ transmit antennas arranged as a uniform linear array (ULA). The radar receiver employs an array of $M_R$ receive antennas with an arbitrary linear configuration. Without loss of generality, a single-element communication receiver is assumed to be located in the direction $\theta_{\rm c}$, which is known to the transmitter.

Let ${\mathbf s}(t) = [s_1 (t), s_2 (t), ..., s_{M_T}(t)]^T \in {\mathbb C}^{M_T \times 1}$ be the baseband equivalent of the signal transmitted by a MIMO radar. Suppose a target is located at $\theta$ with delay $\tau$. The received signal is then given by 
\begin{eqnarray}
{\mathbf r}(t) = \gamma {\mathbf a}_{r} (\theta) {\mathbf a}_{t} (\theta)^T {\mathbf s}(t - \tau) + {\mathbf n}_{\rm R}(t),
\end{eqnarray}
where ${\mathbf a}_t (\theta)$ and ${\mathbf a}_r (\theta)$ are the steering vectors of the transmit and receive array, and $\gamma$ is the coefficient accounting for both target reflection and propogation loss. The radar needs to detect the presence of the target, i.e., $\gamma = 0$ for ${\cal H}_0$ and $\gamma \neq 0$ for ${\cal H}_1$, and estimate the paramters $\theta$ and $\tau$. Assuming a single-antenna communication receiver and considering a sophisticated single-pulse MIMO radar, the baseband signal at the output of the communication receiver can be expressed as
\begin{eqnarray}
y(t) &=& u {\mathbf a}_{t} (\theta_{\rm c})^T {\mathbf s}(t) + n_{\rm C}(t) \nonumber \\
&=& u {\mathbf a}_{t} (\theta_{\rm c})^T \sum_{i} \mathbf{\tilde s}_i(t) + n_{\rm C}(t) ,
\end{eqnarray}
where $u$ is the channel coefficient of the received signal encapsulating the propagation environment between the transmit array and the communication receiver, and $\mathbf{\tilde s}_i(t)$ is the transmitted radar signal in the $i$-th sub-pulse.

The fine structure of the transmitted signal ${\mathbf s}(t)$ dictates the information embedding method. Proposed strategies include:

1) Waveform diversity-based information embedding \cite{blunt2010embedding}. Here $N_b$ bits of information per pulse are embeded by selecting the radar waveform on a pulse-to-pulse basis from a set of $K = 2^{N_b}$ waveforms \cite{ahmad}. Assume that the $k$-th communication symbol is embedded in the $i$-th pulse. Then the corresponding transmit signal vector can be expressed as 
	\begin{eqnarray}
	\mathbf{\tilde s}_i(t) = \sqrt{P_t} {\mathbf 1}_{M_T} \psi_k(t - i T_r),
	\end{eqnarray}
	where $P_t$ is the transmitting power, ${\mathbf 1}_{M_T}$ is the $M_T \times 1$ vector of 1, and $\psi_k(t)$ for $k=1,2,...,K$ are orthogonal waveforms.

2) Phase-modulation based information embedding \cite{sahin2017novel,sahinexperimental,nowak2016mixed}. Information is embedded by controlling the phase of the signal. Assume that the $k$-th communication symbol $b(k)$ is transmitted through the phase information of the constant-envelope vector $\mathbf v = [v(0), v(1),...,v(P_r-1)]^T$. Suppose the symbol $\mathbf v$ is in turn embedded in a single antenna radar waveform, then the total transmit signal is given by 
	\begin{equation}
	\label{eq:PM}
	s(t)=  \sum_{p=0}^{P_r-1} v(p) c(p) \psi(t-p T_r),
	\end{equation}
	where the radar phase modulation $c(p)$ enables direct control of the degree of range sidelobe modulation (RSM). RSM occurs due to the changing waveform structure during the coherent processing interval (CPI) \cite{tian2014novel}, by trading off bit error rate (BER) and/or data throughput. When not properly addressed, RSM translates to residual clutter in the range-Doppler response, and hence degraded target visibility \cite{sahin2017filter,blunt2018radar}: receive filter design to mitigate RSM is addressed for this type of information-embedding in \cite{sahin2017filter,sahin2017novel}. Design methods focus on the realization of a common filter response and exploit the inherent commonality among the radar/communication waveforms. It is worth noting that phase modulation will also inevitably lead to spectrum alteration of the radar waveform, which may result in energy leakage outside the assigned bandwidth \cite{chen2017energy}.
	
3) Sidelobe amplitude modulated-based communications \cite{euziere2014dual,mccormick2017simultaneous,mccormick2017}. To embed the $k$-th communication symbol $b(k)$ during the $i$-th pulse, the beamforming weight vector $\mathbf c_k$ should be associated with that symbol. The amplitude modulated-based method models the transmit signal during the $i$-th pulse as
\begin{eqnarray}
\mathbf{\tilde s}_i(t) = \sqrt{P_t} {\mathbf c}_k \psi(t - i T_r).
\end{eqnarray}
The design of ${\mathbf c}_k$ is formulated as the following optimization problem \cite{hassanien2016dual}:
\begin{eqnarray}
&& \min_{{\mathbf c}_k} \max_{\theta} \left| G(\theta) - |\mathbf c_k^H {\mathbf a}_t(\theta)|  \right|, \theta \in \Theta, \\
&& {\rm s.t.} ~~~ |\mathbf c_k^H {\mathbf a}_t(\theta)| \leq \epsilon, \theta \in \bar\Theta, \mathbf c_k^H {\mathbf a}_t(\theta_{\rm c}) = \Delta_k, \nonumber
\end{eqnarray}
where $G(\theta)$ is the desired transmit beam-pattern, $\Theta$ is the spatial sector the radar keeps under surveillance, $\bar \Theta$ is the sidelobe region for communication, $\epsilon$ is a positive number of user’s choice for controlling the sidelobe levels, and $\Delta_k$ is the $k$-th sidelobe level toward the communication direction $\theta_{\rm c}$. Several other variations of the sidelobe modulating approach are discussed in \cite{hassanien2016efficient,hassanien2016phase}.
	
4) Multi-waveform Amplitude Shift Keying-based information embedding \cite{hassanien2016dual}. This method uses multiple waveforms and two transmit beamforming weight vectors ${\mathbf c}_H$ and ${\mathbf c}_L$. The method requires $N_b$ orthogonal waveforms to embed $N_b$ bits per radar pulse. Then, $N_b$ waveforms are transmitted simultaneously, where the total transmit energy $P_t$ is divided equally among the $N_b$ waveforms. Every transmitted waveform is used to deliver one information bit and the waveform $\psi_k(t)$, $k = 1,2,...,N_b$, is radiated either via ${\mathbf c}_H$ for $b_i(k) = 0$ or ${\mathbf c}_L$ for $b_i(k) = 1$ \cite{ahmad}. The transmit signal is then
\begin{eqnarray}
&& \mathbf{\tilde s}_i (t) = \nonumber \\
&&  \sqrt{\frac{P_t}{N_b}} \sum_{k=1}^{N_b} \left( (1 - b_i(k)) {\mathbf c}_H + b_i(k) {\mathbf c}_L \right)  \psi_k(t - i T_r). 
\end{eqnarray}
	
%\end{itemize}

\color{black}

\subsection{Radar employing	communication waveforms}

Another evolution of functional co-existence is to exploit the waveforms transmitted by a communication network in order to perform sensing (radar) functions. Without loss of generality, we assume a single-element communication transmitter (or a phased-array with an extremely directional beam-pattern). The baseband signal at the communication TX is given by \eqref{eq:xi} with $x_i(t)$ and $v_i(p)$ replaced by $x(t)$ and $v(p)$, respectively. 

Suppose the radar is equipped with $M_R$ antennas and the communication TX is located at angle $\theta_{\rm c}$. There are a number of scattering centers (targets), the $i$-th of which is with path delay $\tau_{i}$, Doppler shift $\nu_i$ and angle $\theta_i$. Let $\gamma_i$ be the coefficient accounting for both target reflection and propagation loss of the $i$-th target. The response from the communication TX to the radar RX in \eqref{eq:rj} can be re-written as 
\[
u_{j}^{\rm CR}(t) = u {a}_{r,j}(\theta_{\rm c}) \delta(t-\tau_{\rm c}) + \sum_i \gamma_i {a}_{r,j}(\theta_i) e^{j 2\pi \nu_i t} \delta(t-\tau_{i}), 
\]
where ${a}_{r,j}(\theta)$ is the angle response of the $j$-th radar RX, $u$ is the coefficient of the direct path between the communication TX and radar RX, and $\tau_{\rm c}$ is the delay of the direct path. As no radar TX is used, the baseband equivalent signal at the radar RX can be obtained from \eqref{eq:rj} with $\sum_{i=1}^{M_T} a_{i,j} s_i(t - \tau_{i,j})$ and $\sum_{i=1}^{M_T} (a_{i,j}^{\rm I} * s_i)(t)$ removed:
\begin{eqnarray}
\label{eq:zt}
{\mathbf r}(t) &=& u {\mathbf a}_r(\theta_{\rm c}) x(t - \tau_{\rm c}) \nonumber\\ 
&& + \sum_i \gamma_i e^{j 2\pi \nu_i t} {\mathbf a}_r(\theta_i) x(t - \tau_{i}) + {\mathbf n}_{\rm R}(t), 
\end{eqnarray}
where ${\mathbf a}_r(\theta) = [{a}_{r,1}(\theta), {a}_{r,2}(\theta), ..., {a}_{r,M_R}(\theta)]^T \in {\mathbb C}^{M_R}$ is the receive steering vector.

One option to use a communication waveform $x(t)$ for sensing is the opportunistic radar based on the 802.11ad standard proposed in \cite{Heath,802.11ad}. The adoption of the 802.11ad standard for 5-th Generation (5G) wireless systems and the exploitation of millimeter Waves (mmWaves) in the 28 and 60 GHz bandwidths \cite{Rappaport_1} immediately raised interest towards the exploitation for sensing applications of some key characteristics of the proposed standard. Indeed, mmWaves suffer from heavy atmospheric attenuation, resonance in the $\text{O}_2$ molecule, absorption by rain, and almost complete shadowing by obstacles, thus requiring Line-of-Sight (LOS) paths between transmitter and receiver. This is in turn achievable thanks to extremely directional beam-patterns and frequent scanning procedures during which the surrounding space is swept in search of nodes willing to establish directional links. As a consequence, the so-called Sector Level Sweep (SLS) phase of the beamforming training protocol provides signals of {\em opportunity} which can be exploited for short-range obstacle detection, typically in {\em automotive} applications \cite{Heath}. In such a phase, the transmitted signal consists of a preamble, containing concatenated complementary Golay codes, and a payload, containing data. The proposed architectures rely on the presence of a receiver, co-located with the wireless transmitter and accessing some key information such as the timing, as well as part if not all of the transmitted signal. With reference to \eqref{eq:zt}, $\tau_{\rm c} = 0$, $u=0$ because there is no direct path, and $x(t)$ is either partially known, since the preamble has a fixed structure, or completely known, if the transmitted data are communicated to the radar receiver. 

Suppose there is one target in each sector. We denote by $\gamma$ its unique complex scattering coefficient. A number of receiving structures have been proposed for target detection and localization in the range/Doppler domain in \cite{Heath,802.11ad}, mostly based on Generalized Likelihood Ratio Test (GLRT) \cite{key1993fundamentals} and assuming different degrees of prior knowledge and cooperation between the radar receiver and the communication transmitter:
\begin{itemize}
	\item[1)] GLRT-1: Everything but the triplet $(\gamma, \nu, \tau)$ in \eqref{eq:zt} is known;
	\item[2)] GLRT-1, simpl.: The receiver is as GLRT-1, but only processes the preamble;
	\item[3)] GLRT-2, SW-1: Like GLRT-1, but $\gamma$ is a nuisance parameter, modeled as complex Gaussian;
	\item[4)] GLRT-3: The payload data are not available to the radar receiver;
	\item[5)] GLRT-4 SW-1: Like GLRT-3, but with $\gamma$ a nuisance parameter;
	\item[6)] Preamble-det: The preamble detector of \cite{Heath}.
\end{itemize}
We underline here that the GLRT strategy is aimed at solving composite hypotheses tests, namely wherein the densities under the two alternatives contain unknown parameters. In practice, these parameters are replaced by the corresponding Maximum-Likelihood (ML) estimates, performed with the same set of data used to make the final decision. Consequently, the GLRT considers, as a by-product, an estimate of the unknown parameters.

\begin{figure}[h]
	\centering
	\includegraphics[width=6cm]{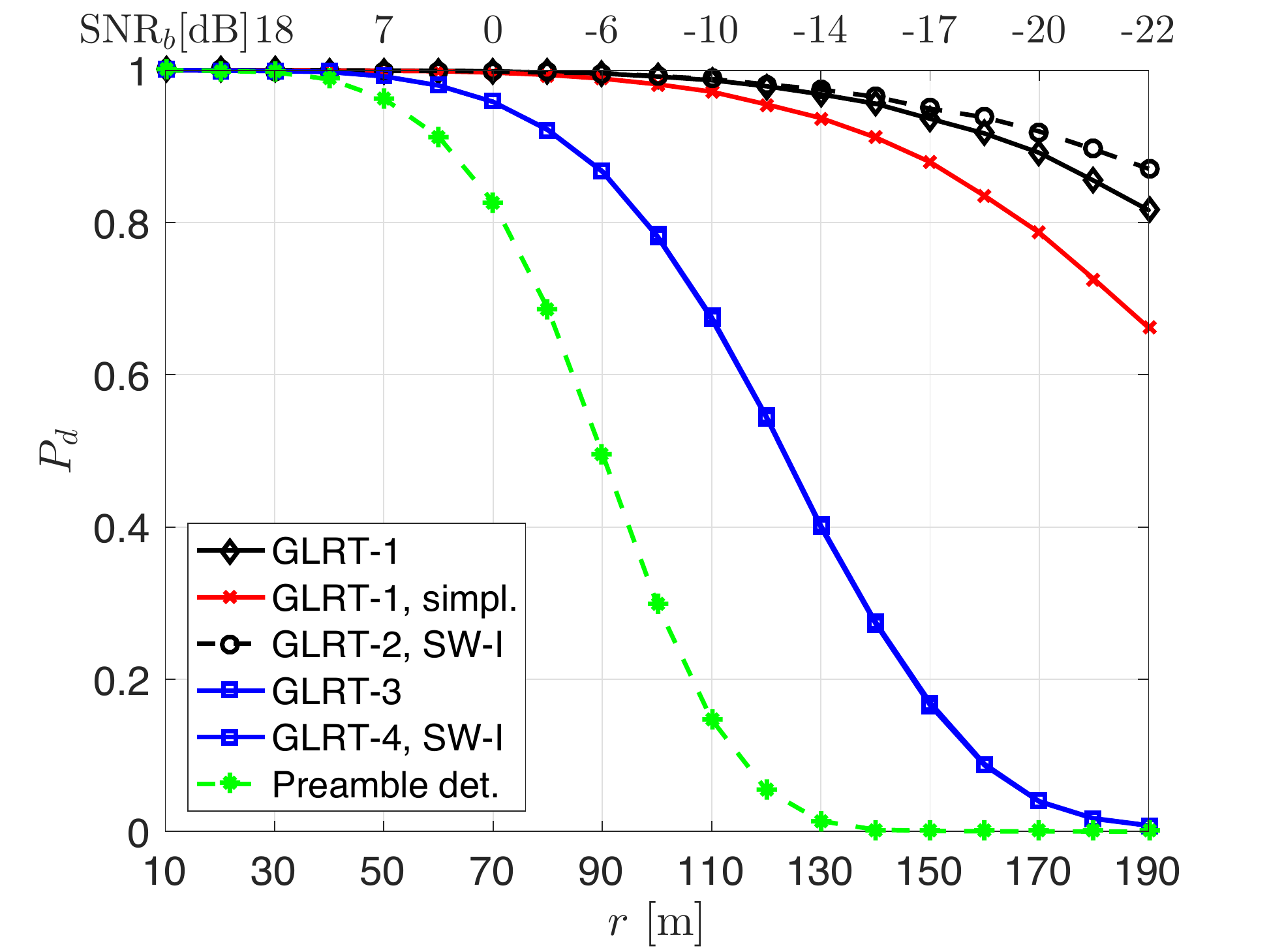}
	\caption{Detection probability as a function of the target range and of the SNR per bit. The false alarm probability is set at $P_{fa}=10^{-4}$. The Figure is taken from \cite{802.11ad}.}
	\label{fig:Pd-mm}
\end{figure}

\begin{figure}[h]
	\centering
	\includegraphics[width=6cm]{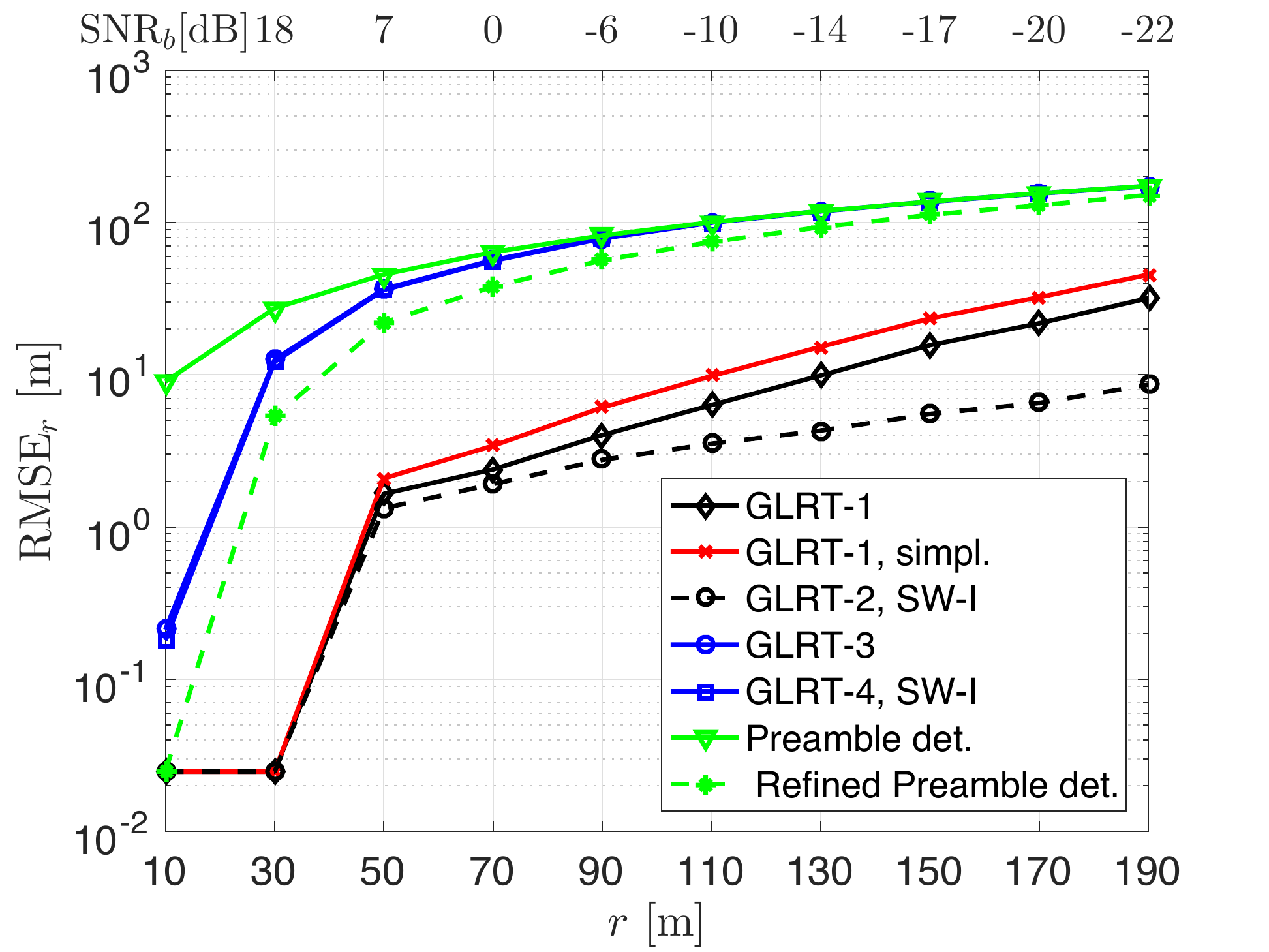}
	\caption{Ranging accuracy as a function of the target range and of the SNR per bit. Figure is taken from \cite{802.11ad}.}
	\label{fig:acc-mm}
\end{figure}

Figures \ref{fig:Pd-mm} and \ref{fig:acc-mm} represent examples of what can be achieved with such opportunistic structures in terms of both detection and localization of an obstacle in short-range applications.
Notwithstanding the encouraging results so far available, a number of problems still remain before claims can be made on the feasibility of such structures. The channel models underlying the results of Figures \ref{fig:Pd-mm} and \ref{fig:acc-mm}, are very simple, assuming that either a single object is present or that it absorbs all of the radiation, thus shielding further obstacles. Moreover, since the range resolution is on the order of decimeters, most objects are typically range-spread, a situation not accounted for so far in the open literature.

Passive radar is another option that exploits other transmissions (communications, broadcast, or radio navigation) rather than having its own dedicated radar transmitter \cite{griffiths2015radar,griffiths2009passive}. It is generally necessary to have a reference channel (RC) dedicated to acquiring the direct path signal as the reference waveform for matched filtering, and surveillance channels (SCs) from which the target reflections are acquired. {For communication TX with known position, $\theta_{\rm c}$ in \eqref{eq:zt} can be obtained \cite{zheng2017super}. The signal in the RC is given by}
\begin{eqnarray}
{z}_{\rm RC}(t) = {\mathbf a}_r(\theta_{\rm c})^H {\mathbf z}(t) = u x(t - \tau_{\rm c}) + {n}_{\rm RC}(t), 
\end{eqnarray}
where ${n}_{\rm RC}(t) = {\mathbf a}_r(\theta_{\rm c})^H \left( \sum_i \gamma_i {\mathbf a}_r(\theta_i) x(t - \tau_{i}) + {\mathbf n}_{\rm R}(t) \right)$. 

{The SC signal is obtained via beamforming on direction $\tilde \theta$:
\begin{eqnarray}
&&{z}_{\rm SC}(t) = {\mathbf a}_r(\tilde \theta)^H {\mathbf z}(t) \nonumber \\
&&= u \zeta_{\rm c} x(t - \tau_{\rm c}) + \sum_i \gamma_i e^{j 2\pi \nu_i t} \zeta_i x(t - \tau_{i}) + {n}_{\rm SC}(t), \nonumber\\
\end{eqnarray}
where $\zeta_{\rm c} = {\mathbf a}_r(\tilde \theta)^H {\mathbf a}_r(\theta_{\rm c})$, $\zeta_{i} = {\mathbf a}_r(\tilde \theta)^H {\mathbf a}_r(\theta_{i})$, and $n_{\rm SC}(t) = {\mathbf a}_r(\tilde \theta)^H {\mathbf n}_{\rm R}(t)$.} To detect the target at delay $\tau$ and Doppler $\nu$, the signal is ``match-filtered" via \cite{berger2010signal}
\begin{eqnarray}
r(\tau) = \int {z}_{\rm SC}(t) e^{-j 2\pi \nu t} {z}_{\rm RC}^*(t - \tau + \tau_{\rm c}) dt.
\end{eqnarray}
The surveillance signal ${z}_{\rm SC}(t)$ contains the signal from the direct path, which causes strong interference. Another issue is that the RC is not very clean in many practical cases, and the performance of the radar is significantly degraded when there is lots of interference, clutter and noise. 

{To improve the performance of passive radar, one can make use of structural information of the underlying communication signal. In particular, since the type of modulation is typically known, we can first estimate the data symbols by demodulation. As demodulation provides better accuracy than directly using the signal in the RC, detection and estimation performance of such radar systems may improve \cite{berger2010signal,zheng2017super}. It is worth noting that passive radar operation is generally inferior to active radar operation due to non-optimal waveforms, spatial beampatterns, and transmit power \cite{blunt2018radar}. Some recent works proposed ``commensal radar" \cite{griffiths2015waveform,ravenscroft2017tandem},  in which the communication signal is designed with the double purpose of transferring information and improving target localization (through a careful autocorrelation function shaping) for a co-existing passive sensing system.}

\section{Conclusions}

We reviewed some of the main ideas and techniques to allow coexistence of sensing and communication functions sharing the same frequency spectrum. The strategies so far proposed have been grouped into three major categories: the first one allows spectral overlap between the signal transmitted by the radar and communication systems, while the other two avoid mutual interference either by cognitively assigning disjoint sub-bands to the different services or allowing just one transmitter to be active and guaranteeing {\em functional} co-existence. For each of the above categories, the basic ideas are outlined, discussing advantages and disadvantages, and offering some examples to illustrate their performance. In the future, hardware prototypes should be built and deployed to be tested on real data. This would permit assessing their performance in real world conditions, including different types of noise, clutter and interference.

%\linespread{1.5}

\bibliographystyle{IEEEtran}
\bibliography{database}

\begin{thebibliography}{10}
\providecommand{\url}[1]{#1}
\csname url@samestyle\endcsname
\providecommand{\newblock}{\relax}
\providecommand{\bibinfo}[2]{#2}
\providecommand{\BIBentrySTDinterwordspacing}{\spaceskip=0pt\relax}
\providecommand{\BIBentryALTinterwordstretchfactor}{4}
\providecommand{\BIBentryALTinterwordspacing}{\spaceskip=\fontdimen2\font plus
\BIBentryALTinterwordstretchfactor\fontdimen3\font minus
  \fontdimen4\font\relax}
\providecommand{\BIBforeignlanguage}[2]{{%
\expandafter\ifx\csname l@#1\endcsname\relax
\typeout{** WARNING: IEEEtran.bst: No hyphenation pattern has been}%
\typeout{** loaded for the language `#1'. Using the pattern for}%
\typeout{** the default language instead.}%
\else
\language=\csname l@#1\endcsname
\fi
#2}}
\providecommand{\BIBdecl}{\relax}
\BIBdecl

\bibitem{griffiths2015radar}
H.~Griffiths, L.~Cohen, S.~Watts, E.~Mokole, C.~Baker, M.~Wicks, and S.~Blunt,
  ``Radar spectrum engineering and management: Technical and regulatory
  issues,'' \emph{Proceedings of the IEEE}, vol. 103, no.~1, pp. 85--102, 2015.

\bibitem{blunt2018radar}
S.~Blunt and E.~Perrins, ``Radar \& communication spectrum sharing,'' 2018.

\bibitem{li2016optimum}
B.~Li, A.~P. Petropulu, and W.~Trappe, ``Optimum co-design for spectrum sharing
  between matrix completion based {MIMO} radars and a {MIMO} communication
  system,'' \emph{IEEE Transactions on Signal Processing}, vol.~64, no.~17, pp.
  4562--4575, 2016.

\bibitem{CME18}
D.~Cohen, K.~V. Mishra, and Y.~C. Eldar, ``Spectrum sharing radar: Coexistence
  via {Xampling},'' \emph{IEEE Trans. Aerosp. Electron. Syst.}, vol.~54, no.~3,
  pp. 1279--1296, 2018.

\bibitem{evans2016shared}
J.~B. Evans, ``Shared spectrum access for radar and communications
  ({SSPARC}),'' \emph{DARPA, Press Release.[Online]. Available: http://www.
  darpa. mil/program/shared-spectrum-access-for-radar-and-communications},
  2016.

\bibitem{patole2017automotive}
S.~M. Patole, M.~Torlak, D.~Wang, and M.~Ali, ``Automotive radars: A review of
  signal processing techniques,'' \emph{IEEE Signal Processing Magazine},
  vol.~34, no.~2, pp. 22--35, 2017.

\bibitem{deng2013interference}
H.~Deng and B.~Himed, ``Interference mitigation processing for spectrum-sharing
  between radar and wireless communications systems,'' \emph{IEEE Transactions
  on Aerospace and Electronic Systems}, vol.~49, no.~3, pp. 1911--1919, 2013.

\bibitem{sanders2012analysis}
F.~H. Sanders, R.~L. Sole, J.~E. Carroll, G.~S. Secrest, and T.~L. Allmon,
  \emph{Analysis and resolution of {RF} interference to radars operating in the
  band 2700-2900 {MHz} from broadband communication transmitters}.\hskip 1em
  plus 0.5em minus 0.4em\relax US Department of Commerce, National
  Telecommunications and Information Administration, 2012.

\bibitem{aubry2014radar}
A.~Aubry, A.~De~Maio, M.~Piezzo, and A.~Farina, ``Radar waveform design in a
  spectrally crowded environment via nonconvex quadratic optimization,''
  \emph{IEEE Transactions on Aerospace and Electronic Systems}, vol.~50, no.~2,
  pp. 1138--1152, 2014.

\bibitem{aubry2015new}
A.~Aubry, A.~De~Maio, Y.~Huang, M.~Piezzo, and A.~Farina, ``A new radar
  waveform design algorithm with improved feasibility for spectral
  coexistence,'' \emph{IEEE Transactions on Aerospace and Electronic Systems},
  vol.~51, no.~2, pp. 1029--1038, 2015.

\bibitem{zheng2018adaptive}
L.~Zheng, M.~Lops, and X.~Wang, ``Adaptive interference removal for
  uncoordinated radar/communication coexistence,'' \emph{IEEE Journal of
  Selected Topics in Signal Processing}, vol.~12, no.~1, pp. 45--60, 2018.

\bibitem{Liu2018MIMO}
F.~Liu, C.~Masouros, A.~Li, T.~Ratnarajah, and J.~Zhou, ``{{MIMO} radar and
  cellular coexistence: A power-efficient approach enabled by interference
  exploitation},'' \emph{IEEE Transactions on Signal Processing}, 2018.

\bibitem{tuninetti}
N.~Nartasilpa, A.~Salim, D.~Tuninetti, and N.~Devroye, ``Communications system
  performance and design in the presence of radar interference,'' \emph{IEEE
  Transactions on Communications}, vol.~XX, no.~X, p. XXX, 2018.

\bibitem{lipetropulu}
B.~Li and A.~P. Petropulu, ``Joint transmit designs for coexistence of {MIMO}
  wireless communications and sparse sensing radars in clutter,'' \emph{IEEE
  Transactions on Aerospace and Electronic Systems}, vol.~53, no.~6, pp.
  2846--2864, 2017.

\bibitem{clancy}
J.~Mahal, A.~Khawar, A.~Abdelhadi, and C.~Clancy, ``{Spectral coexistence of
  {MIMO} radar and {MIMO} cellular systems},'' \emph{IEEE Trans. Aerospace and
  Electronic Systems}, vol.~53, no.~2, pp. 655--668, 2017.

\bibitem{qian2018joint}
J.~Qian, M.~Lops, L.~Zheng, X.~Wang, and Z.~He, ``Joint system design for
  co-existence of {MIMO} radar and {MIMO} communication,'' \emph{IEEE
  Transactions on Signal Processing}, 2018.

\bibitem{zheng2018joint}
L.~Zheng, M.~Lops, X.~Wang, and E.~Grossi, ``Joint design of overlaid
  communication systems and pulsed radars,'' \emph{IEEE Transactions on Signal
  Processing}, vol.~66, no.~1, pp. 139--154, 2018.

\bibitem{khawar2014spectrum}
A.~Khawar, A.~Abdel-Hadi, and T.~C. Clancy, ``Spectrum sharing between {S}-band
  radar and {LTE} cellular system: A spatial approach,'' in \emph{Dynamic
  Spectrum Access Networks (DYSPAN), 2014 IEEE International Symposium
  on}.\hskip 1em plus 0.5em minus 0.4em\relax IEEE, 2014, pp. 7--14.

\bibitem{puglielli2016design}
A.~Puglielli, A.~Townley, G.~LaCaille, V.~Milovanovi{\'c}, P.~Lu,
  K.~Trotskovsky, A.~Whitcombe, N.~Narevsky, G.~Wright, T.~Courtade
  \emph{et~al.}, ``Design of energy-and cost-efficient massive {MIMO} arrays,''
  \emph{Proceedings of the IEEE}, vol. 104, no.~3, pp. 586--606, 2016.

\bibitem{mishali2011xampling}
M.~Mishali, Y.~C. Eldar, O.~Dounaevsky, and E.~Shoshan, ``Xampling: Analog to
  digital at {sub-Nyquist} rates,'' \emph{IET circuits, devices \& systems},
  vol.~5, no.~1, pp. 8--20, 2011.

\bibitem{mishali2010theory}
M.~Mishali and Y.~C. Eldar, ``From theory to practice: {Sub-Nyquist} sampling
  of sparse wideband analog signals,'' \emph{IEEE Journal of Selected Topics in
  Signal Processing}, vol.~4, no.~2, pp. 375--391, 2010.

\bibitem{eldar2015sampling}
Y.~C. Eldar, \emph{Sampling theory: {Beyond} bandlimited systems}.\hskip 1em
  plus 0.5em minus 0.4em\relax Cambridge University Press, 2015.

\bibitem{cohen2018analog}
D.~Cohen, S.~Tsiper, and Y.~C. Eldar, ``{Analog-to-Digital} cognitive radio:
  Sampling, detection, and hardware,'' \emph{IEEE Signal Processing Magazine},
  vol.~35, no.~1, pp. 137--166, 2018.

\bibitem{cohen2018sub}
D.~Cohen and Y.~C. Eldar, ``Sub-nyquist radar systems: Temporal, spectral, and
  spatial compression,'' \emph{IEEE Signal Processing Magazine}, vol.~35,
  no.~6, pp. 35--58, 2018.

\bibitem{blunt}
S.~D. Blunt, P.~Yathan, and J.~Stiles, ``Intrapulse radar-embedded
  communications,'' \emph{IEEE Transactions on Aerospace and Electronic
  Systems}, vol.~46, no.~3, pp. 1185--1200, 2010.

\bibitem{hassanien2016dual}
A.~Hassanien, M.~G. Amin, Y.~D. Zhang, and F.~Ahmad, ``{Dual-Function
  Radar-Communications: Information Embedding Using Sidelobe Control and
  Waveform Diversity.}'' \emph{IEEE Trans. Sig. Proc}, vol.~64, no.~8, pp.
  2168--2181, 2016.

\bibitem{himedembed}
Z.~Geng, R.~Xu, H.~Deng, and B.~Himed, ``{Fusion of radar sensing and wireless
  communications by embedding communication signals into the radar transmit
  waveform},'' \emph{IET Radar, Sonar, Navigation}, vol.~12, no.~6, pp.
  632--630, 2018.

\bibitem{8386661}
F.~Liu, L.~Zhou, C.~Masouros, A.~Li, W.~Luo, and A.~Petropulu, ``Toward
  dual-functional radar-communication systems: Optimal waveform design,''
  \emph{IEEE Transactions on Signal Processing}, vol.~66, no.~16, pp.
  4264--4279, Aug 2018.

\bibitem{Heath}
P.~Kumari, N.~Gonzalez-Prelcic, and R.~W. Heath, ``Investigating the {IEEE}
  802.11ad standard for millimeter wave automotive radar,'' in \emph{Vehicular
  Technology Conference (VTC Fall), 2015 IEEE 82nd}.\hskip 1em plus 0.5em minus
  0.4em\relax IEEE, 2015, pp. 1--5.

\bibitem{802.11ad}
E.~Grossi, M.~Lops, L.~Venturino, and A.~Zappone, ``Opportunistic radar in
  802.11ad networks,'' \emph{IEEE Transactions on Signal Processing}, vol.~66,
  no.~9, pp. 2441--2454, 2018.

\bibitem{griffiths2009passive}
H.~Griffiths, ``Passive bistatic radar and waveform diversity,'' Defence
  Academy of the United Kingdom Shrivenham (United Kingdom), Tech. Rep., 2009.

\bibitem{blunt2016overview}
S.~D. Blunt and E.~L. Mokole, ``Overview of radar waveform diversity,''
  \emph{IEEE Aerospace and Electronic Systems Magazine}, vol.~31, no.~11, pp.
  2--42, 2016.

\bibitem{DeMaioLops}
A.~De~Maio and M.~Lops, ``Design principles of {MIMO} radar detectors,''
  \emph{IEEE Transactions on Aerospace and Electronic Systems}, vol.~43, no.~3,
  pp. 886--898, 2007.

\bibitem{aubry2013knowledge}
A.~Aubry, A.~DeMaio, A.~Farina, and M.~Wicks, ``Knowledge-aided (potentially
  cognitive) transmit signal and receive filter design in signal-dependent
  clutter,'' \emph{IEEE Transactions on Aerospace and Electronic Systems},
  vol.~49, no.~1, pp. 93--117, 2013.

\bibitem{JSTSP}
L.~Zheng, M.~Lops, and X.~Wang, ``{Adaptive Interference Removal for
  Un-coordinated Radar/Communication Co-existence},'' \emph{IEEE Journal of
  Selected Topics in Signal Processing}, vol.~12, no.~1, pp. 1--16, 2018.

\bibitem{DeMaiocognitive}
M.~Piezzo, A.~De~Maio, A.~Aubry, and A.~Farina, ``Cognitive radar waveform
  design for spectral coexistence,'' in \emph{2013 IEEE Radar
  Conference}.\hskip 1em plus 0.5em minus 0.4em\relax IEEE, 2013.

\bibitem{Stoicacognitive}
A.~Aubry, A.~De~Maio, M.~Piezzo, M.~Naghsh, M.~Soltananian, and S.~Petre,
  ``Cognitive radar waveform design for spectral coexistence in
  signal-dependent interference,'' in \emph{2014 IEEE Radar Conference}.\hskip
  1em plus 0.5em minus 0.4em\relax IEEE, 2014, pp. 474--478.

\bibitem{noam2013blind}
Y.~Noam and A.~J. Goldsmith, ``Blind null-space learning for {MIMO} underlay
  cognitive radio with primary user interference adaptation,'' \emph{IEEE
  Transactions on Wireless Communications}, vol.~12, no.~4, pp. 1722--1734,
  2013.

\bibitem{babaei2013practical}
A.~Babaei, W.~H. Tranter, and T.~Bose, ``A practical precoding approach for
  radar/communications spectrum sharing,'' in \emph{Cognitive Radio Oriented
  Wireless Networks (CROWNCOM), 2013 8th International Conference on}.\hskip
  1em plus 0.5em minus 0.4em\relax IEEE, 2013, pp. 13--18.

\bibitem{zhang2008exploiting}
R.~Zhang and Y.-C. Liang, ``Exploiting multi-antennas for opportunistic
  spectrum sharing in cognitive radio networks,'' \emph{IEEE Journal of
  selected topics in signal processing}, vol.~2, no.~1, pp. 88--102, 2008.

\bibitem{zhang2010dynamic}
R.~Zhang, Y.-C. Liang, and S.~Cui, ``Dynamic resource allocation in cognitive
  radio networks,'' \emph{IEEE Signal Processing Magazine}, vol.~27, no.~3, pp.
  102--114, 2010.

\bibitem{cohen2016towards}
D.~Cohen, A.~Dikopoltsev, R.~Ifraimov, and Y.~C. Eldar, ``Towards {sub-Nyquist}
  cognitive radar,'' in \emph{Radar Conference (RadarConf), 2016 IEEE}.\hskip
  1em plus 0.5em minus 0.4em\relax IEEE, 2016, pp. 1--4.

\bibitem{bar2014sub}
O.~Bar-Ilan and Y.~C. Eldar, ``{Sub-Nyquist} radar via {Doppler} focusing,''
  \emph{IEEE Transactions on Signal Processing}, vol.~62, no.~7, pp.
  1796--1811, 2014.

\bibitem{baransky2014sub}
E.~Baransky, G.~Itzhak, N.~Wagner, I.~Shmuel, E.~Shoshan, and Y.~Eldar,
  ``{Sub-Nyquist} radar prototype: Hardware and algorithm,'' \emph{IEEE
  Transactions on Aerospace and Electronic Systems}, vol.~50, no.~2, pp.
  809--822, 2014.

\bibitem{huang2011learning}
J.~Huang, T.~Zhang, and D.~Metaxas, ``Learning with structured sparsity,''
  \emph{Journal of Machine Learning Research}, vol.~12, no. Nov, pp.
  3371--3412, 2011.

\bibitem{frost2012sidelobe}
S.~W. Frost and B.~Rigling, ``Sidelobe predictions for spectrally-disjoint
  radar waveforms,'' in \emph{Radar Conference (RADAR), 2012 IEEE}.\hskip 1em
  plus 0.5em minus 0.4em\relax IEEE, 2012, pp. 0247--0252.

\bibitem{ahmad}
A.~Hassanien, M.~G. Amin, Y.~D. Zhang, and F.~Ahmad, ``Signaling strategies for
  dual-function radar communications: an overview,'' \emph{IEEE Transactions on
  Aerospace and Electronic Systems Magazine}, vol.~31, no.~10, pp. 36--45,
  2016.

\bibitem{blunt2010embedding}
S.~D. Blunt, M.~R. Cook, and J.~Stiles, ``Embedding information into radar
  emissions via waveform implementation,'' in \emph{Waveform Diversity and
  Design Conference (WDD), 2010 International}.\hskip 1em plus 0.5em minus
  0.4em\relax IEEE, 2010, pp. 000\,195--000\,199.

\bibitem{sahin2017novel}
C.~Sahin, J.~Jakabosky, P.~M. McCormick, J.~G. Metcalf, and S.~D. Blunt, ``A
  novel approach for embedding communication symbols into physical radar
  waveforms,'' in \emph{Radar Conference (RadarConf), 2017 IEEE}.\hskip 1em
  plus 0.5em minus 0.4em\relax IEEE, 2017, pp. 1498--1503.

\bibitem{sahinexperimental}
C.~Sahin, J.~G. Metcalf, A.~Kordik, T.~Kendo, and T.~Corigliano, ``Experimental
  validation of phase-attached radar/communication {(PARC)} waveforms: {Radar}
  performance,'' 2018.

\bibitem{nowak2016mixed}
M.~J. Nowak, Z.~Zhang, L.~LoMonte, M.~Wicks, and Z.~Wu, ``Mixed-modulated
  linear frequency modulated radar-communications,'' \emph{IET Radar, Sonar \&
  Navigation}, vol.~11, no.~2, pp. 313--320, 2016.

\bibitem{tian2014novel}
J.~Tian, W.~Cui, and S.~Wu, ``A novel method for parameter estimation of space
  moving targets,'' \emph{IEEE Geoscience and Remote Sensing Letters}, vol.~11,
  no.~2, pp. 389--393, 2014.

\bibitem{sahin2017filter}
C.~Sahin, J.~G. Metcalf, and S.~D. Blunt, ``Filter design to address range
  sidelobe modulation in transmit-encoded radar-embedded communications,'' in
  \emph{Radar Conference (RadarConf), 2017 IEEE}.\hskip 1em plus 0.5em minus
  0.4em\relax IEEE, 2017, pp. 1509--1514.

\bibitem{chen2017energy}
X.~Chen, Z.~Liu, Y.~Liu, and Z.~Wang, ``Energy leakage analysis of the radar
  and communication integrated waveform,'' \emph{IET Signal Processing},
  vol.~12, no.~3, pp. 375--382, 2018.

\bibitem{euziere2014dual}
J.~Euziere, R.~Guinvarc'h, M.~Lesturgie, B.~Uguen, and R.~Gillard, ``Dual
  function radar communication time-modulated array,'' in \emph{Radar
  Conference (Radar), 2014 International}.\hskip 1em plus 0.5em minus
  0.4em\relax IEEE, 2014, pp. 1--4.

\bibitem{mccormick2017simultaneous}
P.~M. McCormick, S.~D. Blunt, and J.~G. Metcalf, ``Simultaneous radar and
  communications emissions from a common aperture, {Part I}: {Theory},'' in
  \emph{Radar Conference (RadarConf), 2017 IEEE}.\hskip 1em plus 0.5em minus
  0.4em\relax IEEE, 2017, pp. 1685--1690.

\bibitem{mccormick2017}
P.~M. McCormick, B.~Ravenscroft, S.~D. Blunt, A.~J. Duly, and J.~G. Metcalf,
  ``Simultaneous radar and communication emissions from a common aperture,
  {Part II: Experimentation},'' in \emph{Radar Conference (RadarConf), 2017
  IEEE}.\hskip 1em plus 0.5em minus 0.4em\relax IEEE, 2017, pp. 1697--1702.

\bibitem{hassanien2016efficient}
A.~Hassanien, M.~G. Amin, Y.~D. Zhang, and F.~Ahmad, ``Efficient sidelobe {ASK}
  based dual-function radar-communications,'' in \emph{Radar Sensor Technology
  XX}, vol. 9829.\hskip 1em plus 0.5em minus 0.4em\relax International Society
  for Optics and Photonics, 2016, p. 98290K.

\bibitem{hassanien2016phase}
------, ``Phase-modulation based dual-function radar-communications,''
  \emph{IET Radar, Sonar \& Navigation}, vol.~10, no.~8, pp. 1411--1421, 2016.

\bibitem{Rappaport_1}
T.~S. Rappaport, S.~Sun, R.~Mayzus, H.~Zhao, Y.~Azar, K.~Wang, G.~N. Wong,
  G.~N. Schulz, Samimi, and F.~Gutierrez, ``Millimeter wave mobile
  communications for 5g cellular: It will work!'' \emph{IEEE Access}, vol.~1,
  pp. 335--349, 2013.

\bibitem{key1993fundamentals}
S.~Kay, ``Fundamentals of statistical signal processing, volume {II}: Detection
  theory,'' 1993.

\bibitem{zheng2017super}
L.~Zheng and X.~Wang, ``Super-resolution delay-doppler estimation for {OFDM}
  passive radar,'' \emph{IEEE Transactions on Signal Processing}, vol.~65,
  no.~9, pp. 2197--2210, 2017.

\bibitem{berger2010signal}
C.~R. Berger, B.~Demissie, J.~Heckenbach, P.~Willett, and S.~Zhou, ``Signal
  processing for passive radar using {OFDM} waveforms,'' \emph{IEEE Journal of
  Selected Topics in Signal Processing}, vol.~4, no.~1, pp. 226--238, 2010.

\bibitem{griffiths2015waveform}
H.~Griffiths, I.~Darwazeh, and M.~Inggs, ``Waveform design for commensal
  radar,'' in \emph{Radar Conference (RadarCon), 2015 IEEE}.\hskip 1em plus
  0.5em minus 0.4em\relax IEEE, 2015, pp. 1456--1460.

\bibitem{ravenscroft2017tandem}
B.~Ravenscroft, P.~M. McCormick, S.~D. Blunt, J.~Jakabosky, and J.~G. Metcalf,
  ``Tandem-hopped {OFDM} communications in spectral gaps of {FM} noise radar,''
  in \emph{Radar Conference (RadarConf), 2017 IEEE}.\hskip 1em plus 0.5em minus
  0.4em\relax IEEE, 2017, pp. 1262--1267.

\end{thebibliography}

%\linespread{1}

\renewenvironment{IEEEbiography}[1]
{\IEEEbiographynophoto{#1}}
{\endIEEEbiographynophoto}

%\vspace{0.5cm}
\begin{IEEEbiography}{Le Zheng}
(M'17) received the B.S. degree in Communication Engineering from Northwestern Polytechnical University (NWPU), Xi'an, China, in 2009 and the Ph.D. degree in Target Detection and Recognition from the Beijing Institute of Technology (BIT), Beijing, China, in 2015. From Feb, 2015 to Feb, 2018, he was a Postdoctoral Researcher in the Electrical Engineering Department of Columbia University, New York, USA.  Now he works at Aptiv (formerly Delphi), CA, USA as a Principal Radar Systems Engineer. His research interests lie in the areas of radar system, statistical signal processing, wireless communication and high-performance hardware for signal processing.
\end{IEEEbiography}

%\vspace{0.3cm}
\begin{IEEEbiography}{Marco Lops}
 is a professor with the Department of Electrical Engineering and Information Technologies at the University "Federico II" of Naples. He obtained his ``Laurea'' and his Ph.D. degrees from ``Federico II'' University (Naples), where he was assistant (1989-1991) and associate (1991-2000) professor. From March 2000 to November 2018 he was a professor at University of Cassino and Southern Lazio and, in 2009-2011, he was also with ENSEEIHT (Toulouse), first as full professor and then as visiting professor. In fall 2008 he was a visiting professor at University of Minnesota and in spring 2009 at Columbia University. He was selected to serve as a Distinguished Lecturer for the Signal processing Society during 2018-2020. His research interests are in detection and estimation, with emphasis on communications and radar signal processing.
\end{IEEEbiography}
%\vspace{0.3cm}

\begin{IEEEbiography}{Yonina C. Eldar} 
	is a Professor in the Department of Electrical Engineering at the Technion - Israel Institute of Technology, Haifa, Israel, where she holds the Edwards Chair in Engineering. She is also an Adjunct Professor at Duke University, a Research Affiliate with the Research Laboratory of Electronics at MIT and was a Visiting Professor at Stanford University, Stanford, CA. She is a member of the Israel Academy of Sciences and Humanities, an IEEE Fellow and a EURASIP Fellow. She has received many awards for excellence in research and teaching, including the IEEE Signal Processing Society Technical Achievement Award, the IEEE/AESS Fred Nathanson Memorial Radar Award, the IEEE Kiyo Tomiyasu Award, the Michael Bruno Memorial Award from the Rothschild Foundation, the Weizmann Prize for Exact Sciences, and the Wolf Foundation Krill Prize for Excellence in Scientific Research. She is the Editor in Chief of Foundations and Trends in Signal Processing, and serves the IEEE on several technical and award committees.
\end{IEEEbiography}
%\vspace{0.3cm}

\begin{IEEEbiography}{Xiaodong Wang} (S'98-M'98-SM'04-F'08) received the Ph.D degree in Electrical Engineering from Princeton University. He is a Professor of  Electrical Engineering at Columbia University in New York. Dr. Wang's research interests fall in the general areas of computing, signal processing and communications,and has published extensively in these areas. Among his publications is a book entitled ``Wireless Communication Systems: Advanced Techniques for Signal Reception'', published by Prentice Hall in 2003.  His current research interests include wireless communications, statistical signal processing, and genomic signal processing. Dr. Wang received the 1999 NSF CAREER Award, the 2001 IEEE Communications Society and Information Theory Society Joint Paper Award, and the 2011 IEEE Communication Society Award for Outstanding Paper on New Communication Topics. He has served as an Associate Editor for the {\em IEEE Transactions on Communications}, the {\em IEEE Transactions on Wireless Communications}, the {\em IEEE Transactions on Signal Processing}, and the {\em IEEE Transactions on Information Theory}. He is a Fellow of the IEEE and listed as an ISI Highly-cited Author.
\end{IEEEbiography}
\end{document}